\RequirePackage{fix-cm}
\documentclass[smallextended]{svjour3}       
\smartqed  
\usepackage{graphicx}
\usepackage{epstopdf}

\begin{document}

\title{Progress in Nuclear Astrophysics of East and Southeast Asia\thanks{This work is supported in part by the National Key Research and Development program (MOST 2016YFA0400501) from
the Ministry of Science and Technology of China, the Strategic Priority Research Program of Chinese Academy of Sciences (No. XDB34020200) and Grants-in-Aid for Scientific Research of JSPS (20K03958, 17K05459).
For Korea: KIH, SA, TSP and DK are supported by the Institute for Basic Science (IBS-R031-D1).
For Malaysia: HAK, NY and NS are supported in part by MOHE FRGS  FP042-2018A and UMRG GPF044B-2018 grants; AAA by IIUM RIGS2016 grant; MB by  FOS-TECT.2019B.04 and FAPESP 2017/05660-0 grants.
For Taiwan: MRW and GG acknowledge supports from the MOST under Grant No.~109-2112-M-001-004 and
the Academia Sinica under project number AS-CDA-109-M11. KCP is supported by the MOST through grants MOST 107-2112-M-007-032-MY3.
KC was supported by the EACOA Fellowship and by MOST under Grant no. MOST 107-2112-M-001-044-MY3.}
}



\author{Azni Abdul Aziz \and
        Nor Sofiah Ahmad         \and
        S. Ahn \and
        Wako Aoki \and
        Muruthujaya Bhuyan \and
        Ke-Jung Chen \and
        Gang Guo \and
        K. I. Hahn \and
        Toshitaka Kajino* \and
        Hasan Abu Kassim \and
        D. Kim \and
        Shigeru Kubono \and
        Motohiko Kusakabe \and
        A. Li \and
        Haining Li \and
        Z.H.Li \and
        W.P.Liu* \and
        Z.W.Liu \and
        Tohru Motobayashi \and
        Kuo-Chuan Pan \and
        T.-S. Park \and
        Jian-Rong Shi \and
        Xiaodong Tang* \and
        W. Wang \and
        Liangjian Wen \and
        Meng-Ru~Wu \and
        Hong-Liang Yan \and
        Norhasliza Yusof
}


\institute{Azni Abdul Aziz \at
            Department of Physics, Kulliyyah of Science, International Islamic University Malaysia, 25200 Kuantan, Pahang Darul Makmur, Malaysia \\
            \email{azniabdulaziz@iium.edu.my}
            \and
            Nor Sofiah Ahmad \at
              Department of Physics, University of Malaya, 50603 Kuala Lumpur, Malaysia \\
            \email{sva190028@siswa.um.edu.my}
           \and
        S. Ahn  \at
            Center for Exotic Nuclear Studies, Institute for Basic Science (IBS), Daejeon 34126, Korea \\
            \email{ahnt@ibs.re.kr}
            \and
        Wako Aoki \at
            National Astronomical Observatory of Japan, 2-21-1 Osawa, Mitaka, Tokyo 181-8588, Japan \\
            \email{aoki.wako@nao.ac.jp}
            \and
        Muruthujaya Bhuyan \at
            Department of Physics, University of Malaya, 50603 Kuala Lumpur, Malaysia \\
            \email{bunuphy@um.edu.my}
            \and
        Ke-Jung Chen \at
            Institute of Astronomy and Astrophysics, Academia Sinica, Taipei, 10617 \\
            \email{kjchen@asiaa.sinica.edu.tw}
            \and
        Gang Guo \at
            Institute of Physics, Academia Sinica, Taipei, 11529 \\
            School of Mathematics and Physics, China University of Geosciences, Wuhan, 430074 \\
            \email{gangg23@gmail.com}
            \and
        K. I. Hahn \at 
            Center for Exotic Nuclear Studies, Institute for Basic Science (IBS), Daejeon 34126, Korea \\
            Department of Science Education, Ewha Womans University, Seoul 03760, Korea \\
            \email{ihahn@ibs.re.kr}
            \and
        Toshitaka Kajino\at 
        National Astronomical Observatory of Japan, 2-21-1 Osawa, Mitaka, Tokyo 181-8588, Japan \\
        Graduate School of Science, The University of Tokyo, Hongo, Bunkyo-ku, Tokyo 11-0033, Japan \\
        School of Physics, and International Research Center for Big-Bang Cosmology and Element Genesis, Beihang University, Beijing 100083 \\
        \email{kajino@buaa.edu.cn}
        \and
        Hasan Abu Kassim \at
        Department of Physics, University of Malaya, 50603 Kuala Lumpur, Malaysia \\
        \email{hasanak@um.edu.my}
        \and
        D. Kim \at
        Center for Exotic Nuclear Studies, Institute for Basic Science(IBS), Daejeon 34126, Korea \\
        \email{dahee@ibs.re.kr}
        \and
        Shigeru Kubono \at
        Center for Nuclear Study, University of Tokyo, 2-1 Hiroswa, Wako, Saitama 351-0198, Japan  \\
        RIKEN Nishina Center, 2-1 Hirosawa, Wako, Saitama 351-0198, Japan \\
        \email{kubono@riken.jp}
        \and
        Motohiko Kusakabe \at
        School of Physics, and International Research Center for Big-Bang Cosmology and Element Genesis, Beihang University, Beijing 100083 \\
        \email{kusakabe@buaa.edu.cn}
        \and
        A. Li \at
        Department of Astronomy, Xiamen University, Xiamen, Fujian 361005 \\
        \email{liang@xmu.edu.cn}
        \and
        Haining Li \at
        CAS Key Laboratory of Optical Astronomy, National Astronomical Observatories, Beijing 100101 \\
        \email{lhn@nao.cas.cn}
        \and
        Z.H.Li \at
        China Institute of Atomic Energy, Beijing, 102413 \\
        \email{zhli@ciae.ac.cn}
        \and
        W.P.Liu \at 
        China Institute of Atomic Energy, Beijing, 102413 \\
        \email{wpliu@ciae.ac.cn}
        \and
        Z.W.Liu \at
        Yunnan Observatories, Chinese Academy of Sciences, Kunming, 650216 \\
        \email{zwliu@ynao.ac.cn}
        \and
        Tohru Motobayashi \at
        RIKEN Nishina Center, 2-1 Hirosawa, Wako, Saitama 351-0198, Japan \\
        \email{motobaya@riken.jp}
        \and
        Kuo-Chuan Pan \at
        Department of Physics, National Tsing Hua University, Hsinchu, 30013 \\
        Institute of Astronomy, National Tsing Hua University, Hsinchu, 30013 \\
        Center for Informatics and Computation in Astronomy, Hsinchu, 30013 \\
        Center for informatics and Computation in Astronomy, National Tsing Hua University, Hsinchu 30013 \\
        \email{kuochuan.pan@gapp.nthu.edu.tw}
        \and
        T.-S. Park \at
        Center for Exotic Nuclear Studies, Institute for Basic Science(IBS), Daejeon 34126, Korea \\
        \email{tspark@ibs.re.kr}
        \and
        Jian-Rong Shi \at
        CAS Key Laboratory of Optical Astronomy, National Astronomical Observatories, Beijing 100101 \\
        School of Astronomy and Space Science, University of Chinese Academy of Sciences, Beijing 100049 \\
        \email{sjr@nao.cas.cn}
        \and
        Xiaodong Tang \at
        Institute of Modern Physics, Chinese Academy of Sciences, Lanzhou, 730000 \\
        Joint department for nuclear physics, Lanzhou University and Institute of Modern Physics, Chinese Academy of Sciences, Lanzhou, 73000 \\
        \email{xtang@impcas.ac.cn}
        \and
        W. Wang \at
        School of Physics and Technology, Wuhan University, Wuhan 430072 \\
        \email{wangwei2017@whu.edu.cn}
        \and
        Liangjian Wen \at
        Institute of High Energy Physics, Chinese Academy of Sciences, Beijing, 100049 \\
        \email{wenlj@ihep.ac.cn}
        \and
        Meng-Ru~Wu \at 
        Institute of Physics, Academia Sinica, Taipei, 11529 \\
        Institute of Astronomy and Astrophysics, Academia Sinica, Taipei, 10617\\
        \email{mwu@gate.sinica.edu.tw}
        \and 
        Hong-Liang Yan \at
        CAS Key Laboratory of Optical Astronomy, National Astronomical Observatories, Beijing 100101 \\
        School of Astronomy and Space Science, University of Chinese Academy of Sciences, Beijing 100049 \\
        \email{hlyan@nao.cas.cn}
        \and
        Norhasliza Yusof \at
        Department of Physics, University of Malaya, 50603 Kuala Lumpur, Malaysia \\
        \email{norhaslizay@um.edu.my}
}

\date{Received: date / Accepted: date}

\maketitle

\begin{abstract}
Nuclear astrophysics is an interdisciplinary research field of nuclear physics and astrophysics, seeking for the answer to a question, how to understand the evolution of the  Universe with the nuclear processes which we learn. We review the research activities of nuclear astrophysics in east and southeast Asia which includes astronomy, experimental and theoretical nuclear physics and astrophysics. Several hot topics such as the Li problems, critical nuclear reactions and properties in stars, properties of dense matter, r-process nucleosynthesis and $\nu$-process nucleosynthesis are chosen and discussed in further details. 
Some future Asian facilities, together with physics perspectives, are introduced. 
\end{abstract}

\section{\label{intro}Introduction}

\begin{figure*}
\includegraphics
[width=1.0\linewidth]
{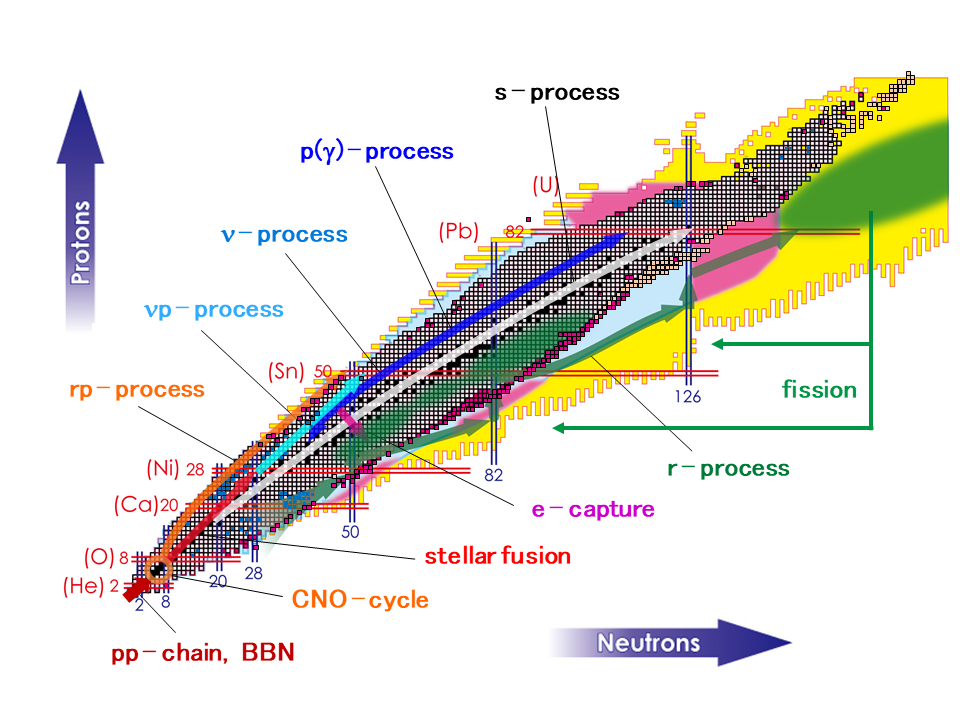}
\caption{\label{fig:nucl_chart} Nuclear chart and the major nucleosynthetic processes in the universe. Most paths involve unstable nuclei, many of which are either difficult
to be studied due to their short lifetimes and low beam intensities, or have not yet been discovered. 226 new isotopes (red boxes) have been synthesized and 84 nuclear masses (blue boxes) have been measured at Asian facilities and adopted in the Atomic Mass Evaluation by 2020~\cite{Wang_2017ChPhCa,Wang_2017ChPhCb}. The nuclei shown in purple-red and light blue indicate those with production rates higher than 1 particle per day, given a primary-beam intensity of 1 particle $\mu$A at RIBF~\cite{Motobayashi2012}. Isotopes in the yellow region are predicted 
based on the KTUY mass formula~\cite{KTUY_mass}
but not synthesized yet. New facilities such as RAON, HIAF 
as well as the RIBF upgrade
will provide new opportunities to synthesize more isotopes which are critical to the understanding of the nuclear processes in stars, galaxies and the universe.}
\end{figure*}

Nuclear astrophysics deals with astronomical phenomena involving atomic nuclei, and therefore it is an interdisciplinary field that consists of astronomy,  astrophysics and nuclear physics. We seek for the answer to a question, how to understand the evolution of the Universe with the nuclear processes which we learn. Figure~\ref{fig:nucl_chart} shows a map of nuclides called nuclear chart, where our knowledge about nuclei and various nucleosynthetic processes are indicated. 

The important questions in our field include but not limited to :
\begin{itemize}
\item What is the origin of the elements in the cosmos?
\item What is the nature of neutron stars and dense nuclear matter?
\item What are the nuclear reactions that drive the evolution of stars and stellar explosions?
\end{itemize}
The research involves close collaboration among researchers in various subfields of astronomy, astrophysics, nuclear and particle physics. This includes astronomical observations using telescopes, gravitational wave (GW) detectors, and neutrino detectors; accelerator laboratory experiments using beams of stable or radioactive nuclei, neutrons, and gamma-rays; laboratory analysis of interstellar grains; large scale computer simulations of stellar explosions and nuclei; and theoretical work in nuclear physics and astrophysics. The interdisciplinary research not only answers the fundamental questions in nature but also provides unprecedented opportunities to discover new physics. For example, the decades of collaborative efforts of astronomers, astrophysicists, nuclear and particle physicists successfully established the standard solar model and led to a perfect solution to the long standing solar neutrino problem and a breaking through discovery of the new properties of neutrino beyond the standard model. A new era of astronomy has been opened by the historic multi-messenger observations of the SN1987A supernova (SN) event and the GW170817 neutron star merger (NSM). Precise knowledge of the critical nuclear physics inputs and reliable stellar models are urgently needed to decipher the messages correctly.

Nuclear astrophysics community in east and southeast Asia has made tremendous progresses over the past few decades. Since early 1980's, the development of radioactive ion beam facilities (RIB) has opened a new frontier of nuclear astrophysics. Short-lived radioactive nuclei are expected to play the critical roles in explosive nucleosynthesis in stars, galaxies and the universe. Such an RIB facility is the laboratory for astrophysics where one can simulate celestial nuclear reactions of astrophysical interest to seek for the origin of atomic nuclides. One of the recent highlights is the  successful approach to the r-process nuclei at RIKEN RI Beam Factory (RIBF). New facilities such as RAON and HIAF are being built to further expand the territory in the unexplored region in the chart of nuclei. The reaction rates of critical reactions, such as the holy grail reaction -- ${}^{12}$C($\alpha$,$\gamma$)$^{16}$O, are the major uncertainties of stellar models, preventing us from understanding the evolution of stars. Jinping Underground Nuclear Astrophysics (JUNA) collaboration developed an underground high-intensity accelerator facility to determine the critical reaction cross sections directly at the Gamow energies, and successfully accomplished the first campaign.  

Remarkable progress in supercomputing power with huge memories enables astrophysicists to simulate catastrophically energetic events like supernova explosions or merging binary neutron stars. We predict a variety of nucleosynthetic products in these explosive episodes as well as quasi-static nuclear burning processes in stars.
These theoretical predictions motivate spectroscopic astronomical observations of elemental abundances and their distribution in the universe through large scale medium-to-high resolution survey including LAMOST-II, 4MOST, WEAVE, etc. as well as high-dispersion spectrographs equipped on board of ground-based large telescopes like Subaru, Keck, VLT or TMT and the space telescope like Hubble Space Telescope.
These astronomy missions are still in middle of the road to reach the oldest metal-free Population III stars that were made from the primeval gas from the
big-bang nucleosynthesis (BBN) in the early universe. 
The ultimate goal is to fully uncover the evolution of the chemical composition in the Universe. Achieving this goal can further bring new insight to the fundamental issues regarding the evolution of dark matter halos.

In this article we first briefly review the nuclear astrophysics programs in five selected countries/regions in east and southeast Asia, and then report the progresses of some hot topics together with some highlights. The future prospect of this field is also provided.

\section{Overview of Asian Activities in Nuclear Astrophysics}
Five countries/regions, namely Japan, Korea, mainland of China, Taiwan and Malaysia, are chosen as the examples to showcase the progress achieved in east and southeast Asia.  

\subsection{Japan}


Nuclear astrophysics in Japan has made remarkable progress since 1980s as a new interdisciplinary field, largely motivated by the development of the studies in nuclear physics with RI beams.

In nuclear physics, extensive studies of neutron halo, which was first discovered by a Japanese group~\cite{Tanihata:1995yv,Tanihata:2013jwa} under INS-LBL (Institute for Nuclear Study - Lawrence Berkeley Laboratory) collaboration, were made at the accelerator facility of RIKEN. Search for new super-heavy elements which led later to the discovery of Nihonium with atomic number Z=113~\cite{Morita-she,Morita2012}, discovery of appearance and disappearance of new and known magic numbers~\cite{Motobayashi2012,Motobayashi-magic}, and measurement of the half lives of r-process nuclei~\cite{LorussoG:2015,Wu:2017,Wu:2020}, respectively, were carried out in this institute. The INS of The University of Tokyo made continuous efforts in developing accelerator technology that would be realized later in HIMAC (Heavy Ion Medical Accelerator in Chiba) and J-PARC (Japan Proton Accelerator Research Complex). 
In particle physics, celestial neutrinos from SN1987A were detected in KAMIOKANDE, 
followed later by discovery of the evidence for neutrino-flavor oscillations in atmospheric neutrinos in SUPER-KAMIOKANDE.
In astronomy, Subaru Telescope had the first light in 1990s, leading to discovery of several extremely metal-poor (EMP) stars which are enriched with r-process elements. In astrophysics, Japanese X-ray satellites HAKUCHO, TENMA, GINGA, ASCA, and SUZAKU identified new X-ray sources in space.

These discoveries were quickly shared and studied theoretically by qualified research scientists crossing over the neighboring interdisciplinary sub-fields (e.g.~\cite{Hashimoto:1989,Thielemann:1990}).  
Such a coherent integration of activities has been supported by the bi-annual international conference 
''Origin of Matter and Evolution of Galaxies'' (OMEG), which started in collaboration with INS (Institute for Nuclear Study) and RIKEN in 1988 and has been held in Asian countries until today. 
Together with OMEG, Japan Forum of Nuclear Astrophysics (JaFNA/UKAKUREN), established in 2008, supports world-wide nuclear astrophysics activities as well as China Institute of Nuclear Astrophysics (CINA), and Joint Institute for Nuclear Astrophysics - Center for the Evolution of the Elements (JINA-CEE).

In late 1980s, the developments of physics of unstable nuclei and the RI beam technology enhanced experimental nuclear astrophysics in Japan along with similar efforts in US and Europe.
An ISOL-based low-energy RI beam facility was sought in INS, while 
RIKEN succeeded in providing in-flight type RI beams in early 1990s. Later RIKEN extended the facility to RIBF, the first "third generation" RIB facility, which started regular operation in 2007 and now approaches r-process nuclei.  The Center for Nuclear Study (CNS) of University of Tokyo introduced a low-energy in-flight RIB facility CRIB (CNS Radioactive Ion Beam separator) in 2000\cite{Kubono_2002}, 
{which was intended to study directly the astrophysical reactions under explosive burning, and has been used to expand international collaborations especially with researchers from the east Asian countries.}
Another large-scale facility J-PARC also provides research opportunities for neutron-induced reactions of  astrophysical interest and spectroscopy of hyper-nuclei which serves to establish the equation-of-state (EoS) of neutron stars.

The first target of the 
experiments in 1980’s was for explosive hydrogen burning, which takes place in novae and X-ray bursts~\cite{Wallace:1981}.   
The $^{15}$O($\alpha,\gamma$)$^{19}$Ne(p,$\gamma$)$^{20}$Na reactions relevant to the breakout of the rp process were studied by the charged-particle spectroscopy~\cite{Kubono:1989}. The onset condition of the hot-CNO cycle was determined by studying the $^{13}$N(p,$\gamma$) reaction by emplying the Coulomb dissociation method~\cite{MotobayashiT:1991}.

Reaction studies by neutron and photon beams were respectively developed by the groups at Tokyo Institute of Technology~\cite{Nagai:1991,Nagai:2020} and Konan University~\cite{Utsunomiya:2003}. A group at Konan University built an experimental hutch called GACKO in the New SUBARU electron ring to study ($\gamma$, n) reactions relevant to p($\gamma$)-process~\cite{Hayakawa:2004} and s-process~\cite{Lugaro:2012,He:2020}. 
In connection with the solar neutrino problem and the overproduction problem of BBN Li abundance, precise theoretical calculations were carried out for both reactions $^3$H$(\alpha,\gamma)^7$Li and $^3$He$(\alpha,\gamma)^7$Be in order to remove nuclear physics uncertainties relevant to these problems~\cite{Kajino1984,Kajino1986}. Also the $^{8}$B production reaction $^{7}$Be(p,$\gamma$)$^{8}$B was investigated by the Coulomb dissociation at RIKEN~\cite{CD8B_RIKEN,Iwasa:1996,Kikuchi:1997} and the resonant scattering of $^{7}$Be+p~\cite{Yamaguchi:2009} at CRIB. An important stellar reaction $^{12}$C($\alpha$,$\gamma$) was intensively studied by a group at Kyushu University by developing a sophisticated recoil mass separator~\cite{Sagara:2005}. 

Japanese astronomers are involved in various observations that provide a direct way to identify the site of nucleosynthesis. 
Recent SN detection is followed up by 
spectroscopic observations with, for example, the Subaru Telescope \cite{Utsumi2017PASJ}, which reveal the various types of SN explosions and identify the elements in their ejecta. Merging binary neutron stars, identified for the first time by the GW detection as GW170817 , could be another site of explosive nucleosynthesis. Follow-up multi-wavelength observations give a significant impact on our understanding of the r-process~\cite{Utsumi_Bulletin2018}. 
Recent observations also revealed that novae are the promising sites of $^7$Li production~\cite{Tajitsu2016ApJ}. 
  
Recent studies with wide-field survey telescopes like Skymapper (e.g.,~\cite{Keller2014Nature}) and large telescopes including the Subaru Telescope (e.g., \cite{Frebel2005Nature,Aoki2014Sci}) found EMP stars having Fe abundances more than 10$^{5}$ times lower than the solar value. Such observations may constrain the mass distribution of first generations of massive stars, which is a long-standing unsolved problem in astronomy. Detailed chemical abundances of metal-poor stars are also applied to constrain r-process, s-process, and the BBN. 

Recently the Gaia mission has precisely measured the position and motion of more than one billion stars.    
Combined with chemical abundance information for a portion of the samples, the formation process of the Milky Way substructures is being studied (e.g., \cite{Matsuno2019ApJ}). Formation of the galactic halo structure should be related to chemical abundances of EMP stars. A  group of stars with large excess of r-process elements (e.g., Eu/Fe ratio more than 10 times higher than the solar value) are of great interest.  The recent discovery of the ultra-faint dwarf galaxy Reticurum II (Ret.~II), whose member stars ubiquitously show enhancement of r-process elements~\cite{Roederer2016AJ,Ji2016Nature}, 
has made a hint to interpret such that these r-process-enhanced stars in the halo could have been formed in a small stellar systems affected by an r-process event, and then accreted into the Milky Way. Recent finding of r-process-enhanced stars in a wide range of metallicity in the Milky Way by a China-Japan collaboration team also shed new light on the interpretation of the the r-process. 

Large progress in these studies is expected in near future. Detection of GW from the neutron star merger event has a large impact on nuclear astrophysics, in particular on the study of r-process. Further progress is expected by new facilities, KAGRA in Japan that started operation in 2020 and LIGO India that is under construction~\cite{Adya:2020}.  
Following the large optical surveys of stars in the Milky Way including those with LAMOST, surveys of fainter objects are planned with the Subaru Telescope using a new instrument Prime Focus Spectrometer (PFS) from 2023. For the follow-up high-resolution spectroscopy for such faint objects, next generation extremely large telescopes are required. China, India, and Japan are the members of the Thirty Meter Telescope (TMT) and Korea and Australia are partners of the Giant Magellan Telescope (GMT). 

\subsection{Korea}
Since the early 1990s, as the nuclear astrophysics community grew along with new advances of Rare Isotope (RI) accelerators around the world, the number of researchers in nuclear astrophysics has steadily increased also in Korea. During the last two decades, many young scientists carried out their researches in nuclear astrophysics at various accelerator facilities such as Center for Nuclear Study (CNS, Japan), Oak Ridge National Laboratory (ORNL, US), National Superconducting Cyclotron Laboratory (NSCL, US), Argonne National Laboratory (ANL, US), TRIUMF (Canada) and RIKEN (Japan). Many of these researchers currently play an important role in the construction of the Korea RI accelerator, RAON, as well as performing experiments at existing RI accelerator facilities around the world.

Measuring the $^{7}$Be(p,$\gamma$)$^{8}$B reaction rate is one of the most important experiments in nuclear astrophysics, because neutrino flux from the $\beta$-decay of $^{8}$B, measured by various underground neutrino detectors to test the standard solar model and the neutrino oscillation, showed a correlation with the reaction rate. The first theoretical calculation on the contribution of s and d states of the $^{7}$Be(p,$\gamma$)$^{8}$B reaction was performed by Korean researchers~\cite{Kim1987}, and motivated a nuclear astrophysics experiment using RI beams, the Coulomb dissociation of $^{8}$B,  investigating the $^{7}$Be(p,$\gamma$)$^{8}$B reaction~\cite{Motobayashi1994}.

While there has been no accelerator producing RI beams in Korea, the Korean nuclear astrophysics community explored many indirect studies of important reactions in astrophysics. Some examples are $^{17}$F(p,$\gamma$)$^{18}$Ne~\cite{Garca1991,Park1999},
$^{19}$Ne(p,$\gamma$)$^{20}$Na~\cite{Smith1992},
$^{7}$Be(p,$\gamma$)$^{8}$B~\cite{Motobayashi1994}, $^{3}$H(p,$\gamma$)$^{4}$He~\cite{Hahn1995},
$^{3}$H(p,n)$^{3}$He~\cite{Brune1999},
$^{13}$N(p,$\gamma$)$^{14}$O~\cite{Smith1993}, $^{14}$O($\alpha$,p)$^{17}$F~\cite{Kim2015},
$^{17}$F($\alpha$,p)$^{20}$Ne~\cite{Cha2017} and $^{22}$Na(p,$\gamma$)$^{23}$Mg~\cite{Kwag2020}, which are related to astrophysical processes including the primordial BBN, the pp chain, the hot CNO cycle, and the breakout reaction to the rp-process.

A close collaboration between Korean and Japanese nuclear astrophysics communities made experimental studies using RI beams at the CRIB. The elastic scattering excitation functions of $^{14,15}$O($\alpha$,$\alpha$), for example, were measured by bombarding radioactive oxygen beams onto the helium gas target to extract the alpha widths and excitation energies~\cite{Kim2015}. Both the important $^{15}$O($\alpha$,$\gamma$)$^{19}$Ne and $^{14}$O($\alpha$,p)$^{17}$F reactions proceed through resonance states above 
the $^{14,15}$O+$\alpha$ thresholds in the explosive process. The $^{26}$Si+p and $^{25}$Al+p elastic scattering experiments~\cite{Jung2012,Jung2014} were also performed to study the $^{26}$Si(p,$\gamma$)$^{27}$P and $^{25}$Al(p,$\gamma$)$^{26}$Si reactions which are well known as the relevant synthesis mechanism of $^{26}$Al$_{gs}$.

The RAON facility~\cite{Jeong2016} is under construction and will be the first RIB facility in Korea, providing exotic and unique radioactive isotopes for key experiments in many areas of nuclear physics and nuclear astrophysics. The stable beam commissioning is expected to begin in 2021 and RI beams will be produced by the Korea Broad acceptance Recoil Spectrometer \& Apparatus (KoBRA)~\cite{Tshoo2016}. Moreover, the Center for Exotic Nuclear Studies, CENS, was recently launched at Institute for Basic Science (IBS). One of the main objectives of CENS is to carry out nuclear astrophysics at existing RIB facilities as well as to prepare future experiments at RAON. 

RAON is also expected to boost active theory-experiment collaborations in nuclear astrophysics. Theory groups in Korea are performing researches on diverse areas to study the nature of nuclear systems and the properties of nuclei. With a chiral perturbation theory, for example, they could resolve the long-standing ~10$\%$ discrepancy between theory and experiment in the radiative neutron-proton capture (n+p$\rightarrow$d+$\gamma$) cross section at threshold~\cite{Park1995}. Another examples are pionless and cluster effective field theories for electro-weak reactions at low-energy, nuclear lattice and no-core shell model for ab initio description of nuclear matter and light nuclei. 

Recently, a new potential dubbed Daejeon16 dedicated to the shell-model and a new energy density functional dubbed KIDS (Korea: IBS-DaeguSKKU) were developed~\cite{Papakonstantinou2018}. Neutrino driven heavy-element synthesis as well as the Li problem in the BBN are also under active investigation.

\subsection{Mainland of China}
The nuclear astrophysics experimental activity in the mainland of China started at China Institute of Atomic Energy (CIAE) in 1990s. The major research facilities for nuclear astrophysics experiments include HI-13 MV tandem accelerator and Beijing Radioactive Isotope Facility (BRIF) -- a ISOL facility driven by proton beam at CIAE, and Heavy Ion Research Facility at Lanzhou (HIRFL) , 320kV accelerator and Low Energy Accelerator Facility (LEAF) at Institute of Modern Physics (IMP), Chinese Academy of Sciences.  The experimental groups have measured a number of critical reaction cross sections and nuclear properties using stable and radioactive ion beams. For example, the CIAE group has performed a number of indirect measurements of important reactions in stars, such as $^7$Be(p,$\gamma$)$^8$B, $^{13}$C($\alpha$,$n$)$^{16}$O and $^{12}$C($\alpha$,$\gamma$)$^{16}$O~\cite{Liu_1996PhRvL,Li_2005PhRvC,Shen_2020PhRvL,Guo_2012ApJ}. The IMP group precisely measured the masses of more than 30 short-lived isotopes in the region close to the proton drip line using a cooler storage ring (CSR) to study the explosive nucleosynthesis~\cite{Xing_2018PhLB..781..358X,Yan_2013ApJ...766L...8Y,Tu_2011PhRvL.106k2501T} and also performed a number of experiments to determine the critical reaction rates using the low energy stable or radioactive beams provided by HIRFL and the CRIB radioactive beam facility at Japan \cite{Zhang_2014PhRvC_18neap,He_2011EPJA,He_2013PhLB}. 
The IMP group is also in charge of the atomic mass evaluation (AME), an essential tool for the studies of nuclear physics and nuclear astrophysics \cite{Wang_2017ChPhCa,Wang_2017ChPhCb}. The Jinping Underground Nuclear Astrophysics  (JUNA) collaboration~\cite{JUNA_2016SCPMA} established an underground accelerator facility for the direct measurements of critical nuclear reaction cross sections at stellar energies. The newly finished back-n white neutron facility at China Spallation Neutron Source (CSNS)~\cite{CSNS_2017JInst} and the to-be-finished Shanghai Laser Electron Gamma Source~(SLEGS) will  offer new opportunities for the studies of neutron or gamma-induced reactions. High Intensity heavy ion Accelerator Facility (HIAF) are being built~\cite{Zhou2019_HIAF} to measure the critical reactions at stellar energies and study the critical nuclear properties. More details of JUNA and HIAF are available in section~\ref{sec:future}. Beijing ISOL, a powerful ISOL facility driven by a reactor, is being proposed~\cite{Liu_2011SCPMA,Cui_2013NIMPB,asia_nuclear_physics_2020} for the studies of the r-process nucleosynthesis. 

Theoretical nuclear model calculations provide an indispensable approach in the region beyond the current experimental capability. For example, a database of neutrino spectra of the sd-shell nuclei under astrophysical conditions was established with the nuclear shell model~\cite{Misch_2018}.  The spherical relativistic continuum Hartree-Bogoliubov (RCHB) theory and mass formula were applied to predict the nuclear masses in the neutron-rich region which is inaccessible at any existing facilities~\cite{Xia_2018_mass,Wang_2014}. The nuclear beta decay half-lives were systematically studied by fully self-consistent relativistic quasiparticle random phase approximation (QRPA) model~\cite{Niu_2013_beta_decay}, and the description of beta-decay half-lives was further improved by either considering tensor force or including many-body correlations beyond~QRPA\cite{Minato_2013_beta_decay,Niu_2018_beta_decay}. New sets of EoS, for example, Quark Mean Field~\cite{LI2020}, have been proposed for neutron stars through the confrontation of theoretical calculations with laboratory measurements of nuclear properties and reactions and increasingly accurate astronomy observations. 

The Large Sky Area Multi-Object Fiber Spectroscopic Telescope (LAMOST) is the largest Chinese optical telescope constructed in 2008.
LAMOST survey \cite{Cui2012,Zhao2012} has finished five-year phase-I survey with the low-resolution mode ($R \sim 1,800$).
Currently, it is running phase-II survey in a combination mode with both low- and medium-resolution spectra ($R \sim 7,500$).
The LAMOST survey has obtained over 10 million low-resolution spectra and six million medium-resolution spectra.
In the 7th data release (DR7) of LAMOST, seven million stars are released for public use with reliable stellar parameters calculated through an upgraded pipeline~\cite{Luo2015}.
Based on the LAMOST data and follow-up observations with other facilities including the Subaru telescope, Lijiang 2.4m and 1.8m telescopes, etc.,
great advance has already been made in the field of stellar and Galactic astronomy.
The huge database of LAMOST has become an invaluable reservoir of studying rare objects leading to a number of important discoveries.
For example, the most Li-rich K giant TYC\,429-2097-1 has been discovered to have a Li abundance over 3,000 times higher than the normal amount
and provides a great opportunity to investigate the origin and evolution of Li in the Galaxy \cite{Yan2018NatAs},
while Yan et al.\cite{Yan2020NatAs} presents a uniquely large systematic study of lithium-rich stars and reveals that the distribution of lithium-rich red giants
declines steeply with increasing lithium abundance, with an upper limit of around 2.6 dex, whereas the lithium abundances of red clump (RC) stars extend to much higher values;
Xing et al.\cite{Xing2019NatAs} reports the first extremely r-process enhanced (r-II) star exhibiting very low $\alpha$-elemental abundance,
which provides the clearest chemical signature of past accretion events onto the Milky Way;
Liu et al.\cite{Liu2019Nature} discovers the most massive stellar black hole whose formation poses an extremely challenging problem to current stellar evolution theories.
On the other hand, LAMOST also enables systematic studies on large sample of various types of stars in the Galaxy,
e.g., the largest metal-poor star searching project, which has discovered over 10,000 candidates with metallicities less than 1\% of the sun \cite{Li2018ApJS},
thus forming the largest catalogue of bright source of metal-poor stars in the world
which is very suitable for statistical studies of the Galactic halo and follow-up observation with the existing ground-based telescopes as well.
LAMOST is now moving forward to another step to install a high-resolution spectrograph.
Though still in the commissioning stage, it has already obtained scientific progresses such as discovering two new Li-rich RC stars \cite{Zhou2020RAA}.
Besides the conventional optical observation, the observation of INSIGHT--hard X-ray modulation telescope helped to confirm the unexpected weak and soft nature of GRB170817A. By collaborating with the MPIE group,
astronomers also explore the $\gamma$-rays from radioactive isotopes in the Milky Way to probe the supernova progenitors and later evolution of massive stars. A 20kt neutrino observatory (JUNO) will be ready in 2020 to
detect all six flavors of neutrinos and anti-neutrinos
emitted from the supernova core~\cite{An:2015jdp,Lu:2016ipr,Li:2017dbg,Li:2019qxi,Li:2020gaz}.

The stellar modeling program is being developed in the mainland of China. The groups at Yunnan observatories have been working on the progenitors of SNe Ia since 2006. Their detailed 1D stellar evolution calculations suggested that SNe Ia generated from Si-rich WDs and/or C-O-Si WDs are expected to show some particular characteristic features such as the high velocity of Si lines in their spectra. This can be used as the 'smoking gun' for identifying the SN Ia progenitor systems composed of a CO WD and a He star~\cite{Wu2020}. They are also working on multi-dimensional hydrodynamical simulations of SNe Ia such as the interaction of SN Ia ejecta with a stellar companion star with a hope to predict observables which can be used to compare with the SN Ia observations with different telescopes~\cite{LiuZW2012}. This is expected to place strict constraints on SN Ia progenitor models and their explosion mechanism.

\subsection{Taiwan}

Earlier efforts in nuclear astrophysics focused on measurements of short-lived radioactive nuclei in meteorites and understanding their origin~\cite{2001ApJ...548.1029S,2001ApJ...548.1051G,SCHONBACHLER2003467,2003ApJ...596L.109S,2006ApJ...640.1163G,2010ApJ...719L..99L,2011ApJ...743L..23C,2015ApJ...806L..21C}, as well as on experimental side relevant to the nuclear reaction rates in astrophysical environments, e.g.,~\cite{Wang:1974zza,1984NuPhA.422..373W,Trentalange:1988rme,2002JPhG...28L...9H}. 
Since 2018, researches focusing on the theoretical aspects of supernova explosions and the nucleosynthesis of heavy elements have become active in various leading institutes, including the Institute of Physics, Institute of Astronomy and Astrophysics at the Academia Sinica, and the Institute of Astronomy at the National Tsing Hua University. 
Moreover, as nuclear astrophysics plays a pivotal role in various phenomena that are the prime targets in the newly-opened era of multi-messenger astronomy, collaborative efforts among domestic researchers are under development supported by the National Center for Theoretical Sciences. 

As for supernova-related studies, efforts in implementing the efficient neutrino transport scheme -- Isotropic Diffusion Source Approximation (IDSA) -- were made in multidimensional simulations of core-collapse supernova explosions with the open-source code FLASH~\cite{2016ApJ...817...72P,2017nuco.confb0703P,2018A&A...619A.118C,2019JPhG...46a4001P}. With such development, the dependence of the black-hole forming supernovae on different nuclear EoS and the stellar rotation rates was examined and the expected GW signals were computed~\cite{2018ApJ...857...13P,2020arXiv201002453P}.
Guo, Wu and their collaborators then improved the treatment and consistency of the neutrino interactions in dense matter, including the inverse neutron-decay, muonic charged-current interactions, and nucleon-nucleon Bremsstrahlung~\cite{Fischer:2018kdt,Guo:2019cvs,Guo:2020tgx,Fischer:2020vie}, which allows to quantify the impact of these interactions on supernova simulations.
Meanwhile, in a series of work~\cite{Fischer:2017lag,Fischer:2020xjl}, potential explosions of massive stars of $\sim 35-50$~$M_\odot$ triggered by the hadron-quark phase transition were suggested. Such exotic explosions were found to result in unique signatures in both the neutrino signals and the $r$-process nucleosynthesis yields that can be tested by future neutrino detection and metal-poor star observations.
Chen et al.~\cite{2020ApJ...897..152C,2020ApJ...893...99C,2019arXiv190412873C} performed detailed multidimensional long-term simulations of the pair-instability supernovae and magnetar-driven supernovae on even more massive and energetic end. These simulations aim to study the important consequences due to multidimensional effects on the observables. 

Regarding the $r$-process nucleosynthesis in NSMs, various unique signatures of the $r$-process, including the late-time kilonova lightcurves and the $\gamma$-ray emission from merger remnants residing in the Milky Way, were explored~\cite{Wu:2018mvg,Wu:2019xrq,Giuliani:2019oot}. Future detection of these can shed new lights on our understanding of the nature of $r$-process in NSMs.
Banerjee, Wu and Yuan examined the important question of whether NSMs can be the dominant $r$-process sites in the Milky Way~\cite{Banerjee:2020eak}, which pointed out that the late-time galactic chemical evolution of $r$-process enrichment can be consistently accounted for by NSMs when considering new effects of neutron star kicks and the inside-out evolution of the Milky Way which was omitted in previous studies and easing the need of additional $r$-process sources.
In parallel, several authors made efforts in studying the onset conditions of neutrino flavor conversions in NSM remnants and their impact on $r$-process\cite{Wu:2017qpc,Wu:2017drk,George:2020veu}.
Improved numerical modeling of this phenomenon is on-going to further solidify the roles of neutrinos in NSMs.

\subsection{Malaysia}
Nuclear astrophysics and astrophysics as a whole is in its infancy stage of developing into a serious discipline in Malaysia where it started in earnest in the early 1990s on the standard solar model and solar neutrinos. The research focuses on the theoretical aspects of nuclear astrophysics. Currently, University of Malaya and International
Islamic University Malaysia are involved in this field.

Amongst the topics of interest in nuclear astrophysics is the study of thermonuclear reaction rates in stellar evolution \cite{yusof2010charged} and on the reliability of the M3Y  double  folding  potentials  (DFM)  and  the $\alpha$-$\alpha$ double-folding cluster (DFC) potentials  in  fusion  cross  sections  of  light  systems  of  heavy ion reactions. Both potentials have been shown to reproduce  the experimental fusion  cross  section data   reasonably  well \cite{aziz2015reliability}.  In fact, the DFC potential reproduces satisfactory fitting to the astrophysical S-factor data and consistent prediction of the astrophysical reaction rates.

In the light of observations of very massive stars (VMS) in the local universe \cite{crowther2010r136}, theoretical models were constructed to verify this observation \cite{yusof2013evolution}. The focus of this work is mainly on the study of life and death of  very massive stars, which are extreme cosmic engines, and enriching their environments with chemically processed material throughout their entire life-time in Ref.\cite{yusof2013evolution,whalen2014pair,gilmer2017pair}. 
This work is in collaboration with Keele Astrophysics and Geneva stellar 
evolution group where both research groups are actively 
involved in the study of the evolution of massive stars. The current project is a study on the grids
of stars with supersolar metallicity which would be useful for  studying massive stars in the metal rich environment.

In the area of  neutrino astrophysics, the stopping power of matter on neutrinos with oscillation effects is applied to cosmological neutrinos \cite{ibrahim2009neutrino} and neutrinos in very massive stars. Recently, neutrino emissivity from thermal processes in VMS in our local universe is investigated. Neutrino luminosity towards  the end of the VMS life is higher than in massive stars. This could shed light on the possibility of using the detection of the neutrinos to locate the candidates for the pair instability supernova in our local universe.

The first detection of binary neutron star merger event GW170817 in 2017 by LIGO and Virgo \cite{abbott2018gw170817} has established limits both on the tidal deformability of the canonical star to be constrained in the range of $70 \leq\Lambda_{1.4} \leq580$, and the chirp mass ${\cal M} = 1.188 M_\odot$ \cite{abbott2018gw170817}. The new constraints from binary neutron star merger in terms of tidal polarization open the possibility to revisit the well-established model, namely, density-dependent van der Waal model, Skyrme-Hartree-Fock, momentum dependence interaction, and relativistic mean field approaches. Further it is crucial to establish the possible correlations and sensitivity of important nuclear bulk properties, i.e., the symmetry energy, its slope, compressibility, and values of the tidal deformability for the canonical $1.4M_\odot$ \cite{lourencco2019density,lourencco2019neutron,lourencco2020consistent,lourencco2020gw170817}. A connection has been established  between the finite nuclei and infinite nuclear matter in terms of isospin asymmetry parameter i.e., symmetry energy and its coefficients using relativistic mean-field formalism \cite{bhuyan2018surface,biswal2020nuclear}. This result is of considerable importance since due to shell closure over the isotonic chain, will act as an awaiting point in nucleosynthesis of r-process and experimental investigations towards the exotic region of the nuclear chart including the superheavy island.

\section{SPECIAL TOPICS}
Here we discuss several hot topics that exhibit a remarkable progress in nuclear astrophysics on synergy of the three fields of astronomy,  nuclear and particle physics, and astrophysics.

\subsection{Li problems}

The Li production in BBN is a longstanding unsolved problem~\cite{bertulani2016}. The standard homogeneous BBN model with cosmological baryon-to-photon ratio $\eta$ determined by the observations of Cosmic Microwave Background (CMB) anisotropies and precisely measured neutron half-life predicts about a factor of three larger Li abundance than those measured in metal-poor main-sequence turn-off stars, while the other light nuclides $^{2}$H, $^{3}$He and $^{4}$He are in reasonable agreement with observational constraints. The discrepancy of $^{7}$Li is remarkably larger than error sources including non-LTE and 3D effect of stellar atmospheres in calculation of spectral line formation, and is called ''Big-bang Li problem'' Recent observations extended to the lowest metallicity range ([Fe/H]$<-3$) suggest systematically lower Li abundances than those at [Fe/H]$>-3$~\cite{Matsuno2017AJ}. This is a stringent constraint on the mechanism that causes the Li problem. 

A number of suggestions have ever been proposed to solve this problem. One is to understand better the effects of diffusive transport and stellar mass-dependent convection and depletion of Li near the stellar surface of the metal-poor halo stars~\cite{Korn2006Natur,Melendez2012AA}. However these astration effects are hard to be fully understood.

This problem has been extensively investigated in nuclear physics laboratories. Although the production reactions $^3$H$(\alpha,\gamma)^7$Li and $^3$He$(\alpha,\gamma)^7$Be(e$^-$,$\nu_e$)$^7$Li are reasonably well studied both experimentally~\cite{Adelberger2011,Vorabbi2019} and theoretically~\cite{Kajino1984,Kajino1986,Neff2011}, there are some reactions not well studied yet which might alter the $^{7}$Li abundance. 
The CIAE group has determined the cross sections of 12 reactions involving lithium \cite{Li_2011_li_problem}. But the BBN calculations showed that the new reaction rates have minimal effect on the final lithium abundances. In fact, 
$^{7}$Be is much more abundantly produced than $^{7}$Li in the standard BBN with the current accepted $\eta$-value. Therefore, the main focus has been placed on the experimental studies of destruction reactions of $^{7}$Be, i.e. $^{7}$Be(n,p)$^{7}$Li and $^{7}$Be(n,$\alpha$)$^{4}$He. 
The JAEA experiment~\cite{Ishikawa:2020} of the former reaction reported significant contributions of the decay brunches to the first excited state of $^{7}$Li from the relevant $^{8}$Be resonances, which was not identified in the previous n-ToF experiment at CERN~\cite{Damone:2018}. Hayakawa et al.~\cite{Hayakawa:2019} have successfully measured both transitions, for the first time, at the wide energy range relevant to BBN, applying Trojan Horse Method (THM) by the use of intense $^{7}$Be beam from CRIB at CNS. They also deduced the reaction cross sections for $^{7}$Be(n,$\alpha$)$^{4}$He as well. 
This reaction was studied at RCNP~\cite{Kawabata:2017} by the use of time reverse reaction $^{4}$He($\alpha$,n)$^{7}$Be and the p-wave dominance was found. These precise experimental data proved that both destruction reactions of $^{7}$Be are not responsible for the factor of three overproduction in $^{7}$Li abundance.

The $^{7}$Be(d,p)$^{8}$Be and $^{7}$Be(d,$^3$He)$^{6}$Li destruction reactions were also  studied at RCNP and CIAE, respectively. Their contributions were found to be quite minor~\cite{Inoue:2018,Li_2018_Li_problem}. An important report from Florida group~\cite{Rijal:2019} is that the $^{7}$Be(d,$\alpha$) reaction might make a remarkable effect on destruction of $^{7}$Be. All other reactions however need to be revisited before making a definite conclusion.

Many other models beyond the standard BBN have also been proposed such as those including non-extensive Tsallis statistics~\cite{Hou:2017uap}, baryon inhomogeneous BBN~\cite{Alcock1987,Kajino1991,Orito:1997}, supersymmetric particle decay~\cite{Arbey2008,Kusakabe:2014moa}, sterile neutrinos~\cite{Esposito2000}, and so on. Among these, Tsallis statistics is one of the interesting  models to solve the big-bang Li problem, and its theoretical establishment has started by taking account of the inhomogeneous primordial magnetic fields~\cite{Luo:2018nth} still satisfying the observed constraints from CMB~\cite{Yamazaki:2012}. 

\begin{figure}
\includegraphics
[width=0.9\linewidth]
{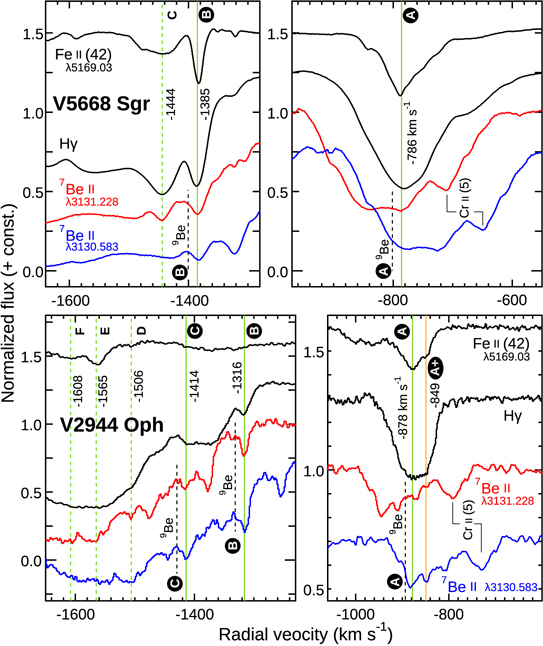}
\caption{\label{fig:7be} $^7$Be spectra detected in novae explosions~\cite{Tajitsu2016ApJ}.}
\end{figure}

Post-cosmological origin of Li has been studied theoretically in chemical evolution models~\cite{Prantzos2012A&A}. 
Tajitsu et al.~\cite{Tajitsu2015Natur} have reported the detection of $^{7}$Be, which decays to $^{7}$Li in 53 days, in the ejecta of a nova explosion (Fig.~\ref{fig:7be}). This first discovery was followed by further detection of $^{7}$Be in other novae~\cite{Tajitsu2016ApJ,Molaro2016} in addition to detection of $^{7}$Li~\cite{2015ApJ...808L..14I}. These observations indicate that nova explosions could be the major production sites of $^{7}$Li in the present Milky Way as predicted theoretically. 

Another unsolved question in astronomy is extremely enhanced Li abundances as high as A(Li) = log($n_{Li}$/n$_{H}$) + 12 $>$ 1.5 in $\sim 1\%$ of low-mass giant stars (e.g.~\cite{Casey2019ApJ}). $^7$Be is produced by the $^3$He$(\alpha,\gamma)^7$Be reaction in the hydrogen burning shell, conveyed to the stellar surface, and eventually decays to $^{7}$Li, which is called Camelon-Fowler (CF) mechanism~\cite{Cameron1971ApJ}. However, since $^7$Li is easily destroyed by the $^7$Li(p,$\alpha$) reaction at low temperature, such a high $^7$Li abundance level is not explained in the standard theory of low-mass stellar evolution. Recent studies have revealed that most Li-rich objects are found in the core He-burning phase (clump stars)~\cite{Yan2020NatAs,Kumar2020NatAs}.
There is a peculiar scenario such that extra mixing is induced by the tidal interaction with companion star and drives Li production. Another scenario is to assume that substellar objects holding high Li abundance were engulfed to form the stellar atmosphere of these stars. It has recently been pointed out that the produced $^7$Be is boosted to the stellar surface due to more efficient themohaline mixing if neutrinos had finite magnetic moment to accelerate stellar cooling~\cite{Mori:2020qqd}. These proposed mechanisms of Li enhancement are still under debate. 

\subsection{\label{Topics:h}Critical nuclear reactions and properties during quiescent and explosive burning}
Critical reactions in stars are essential parts of the stellar models to understand multi-messages from stars, such as the isotopic/elemental abundances, light/neutrino fluxes or even GW of stars. However, the determination of the very low nuclear reaction rates in stars, and the determination of properties and reactions of very neutron deficient or very neutron rich unstable nuclei are two long standing challenges in nuclear astrophysics~\cite{Schatz_2016JPhG}. We review some progresses in the precise measurements of the critical  nuclear reactions and properties. The r-process nucleosynthesis and the weak interaction process are discussed separately in sec.\ref{Topics:r} and sec.\ref{subsec:weak}.  


\subsubsection{Helium and carbon burnings}
The $^{12}$C($\alpha,\gamma$)$^{16}$O reaction is the key reaction in the stellar helium burning stage. Competing with 3$\alpha$ process, the $^{12}$C($\alpha,\gamma$)$^{16}$O reaction rate determines the ratio of carbon and oxygen in the universe and affects many other stellar scenarios such as the synthesis of $^{60}$Fe and $^{26}$Al, supernova explosion and the formation of back holes~\cite{Farmer_2019_black_hole}. $^{12}$C($\alpha,\gamma$)$^{16}$O is quoted as the Holy Grail reaction and a reaction rate with accuracy better than 10\%  is required~\cite{Weaver_1993} .

\begin{figure}[htbp]
\centering
\includegraphics[width=0.9\linewidth]{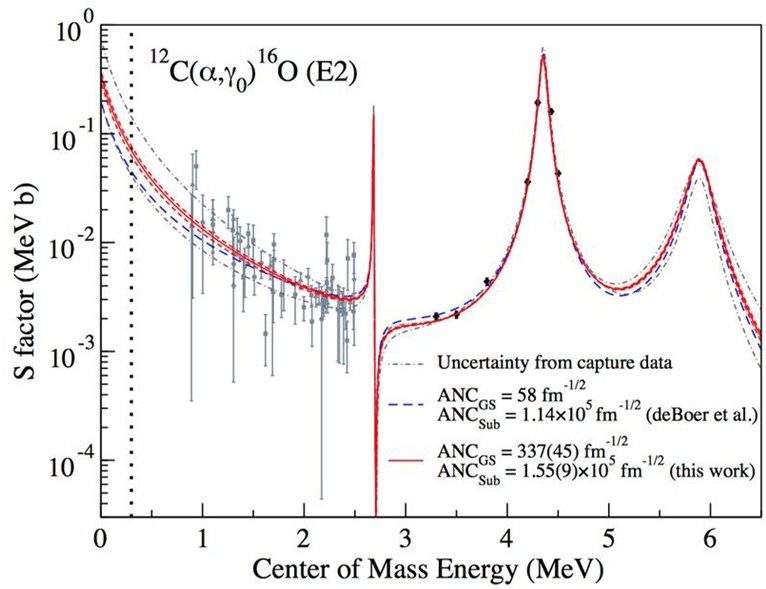}
\caption{The S$_{E2}$(E) factor of the $^{12}$C($\alpha,\gamma$)$^{16}$O reaction. The grey line represents the result constrained by the data of direct measurements. The red line represents the result determined with the ANC of $^{16}$O ground state.}
\label{fig:c12ag}
\end{figure}

This reaction has been studied directly in the ground level labs using normal and inverse kinematics, respectively, in Japan~\cite{Sagara:2005,Makii_c12ag_PhysRevC}. However, the needed cross sections at around 300 keV is far yet to be studied in the coming years~\cite{Makii_c12ag_PhysRevC}. The CIAE group performed an indirect measurement of the S$_{E2}$(300) factor. Comparing to the value constrained by the direct measurement, the uncertainty of S$_{E2}$(300) factor is reduced from 56\% to 10\% , as shown in Fig.~\ref{fig:c12ag}. The Kyushu group initiated a new program to revisit the nuclear reactions of the He burning using the quality beam from a Tandem accelerator.  JUNA collaboration began the direct measurement in the Jinping Underground Laboratory in early 2021(See section~\ref{subsec:juna}). Another direct measurement of the reaction using an active target TPC is proposed in Korea as well~\cite{Ahn2020}.

Important progresses have been made for the triple alpha process in the last decade.
Although a non-resonant mechanism was suggested to dominate the process with CDCC method~\cite{Ogata:2009}, the detailed rate has been derived theoretically~\cite{Akahori:2015}, which turns out to be inefficient to change the evolution of low-to-intermediate mass stars~\cite{Suda:2011}. The triple alpha process at high temperatures are crucial to explosive burning, such as XRB and supernova explosion~\cite{Beard_2017PhRvL}. Resonances of 0$^{+}$ and 2$^{+}$ were observed at around 10 MeV~\cite{Ito:2011} at RCNP. The 2$^{+}$ resonance, which is likely a rotational member of the Hoyle state, was confirmed~\cite{Zimmerman:2013}.
Recently, this width of the 9.6 MeV 3$^-$ resonance was studied experimentally at RCNP, determining the total decay width for the first time. Their preliminary results shows a reduction of the reaction rate.

The ${}^{12}$C+${}^{12}$C fusion reaction is the primary reaction in
the hydrostatic and explosive C/Ne burnings in massive stars. It is also important for the stellar explosions such as Type Ia supernova and superburst~\cite{Gasques2007,Mori_2018}. Determination of the  ${}^{12}$C+${}^{12}$C fusion cross section at stellar energies below E$_{c.m.}$ = 3~MeV has been a long standing problem since the first measurement at 1960. While the $^{12}$C($^{12}$C,$n$)$^{23}$Mg reaction has been measured directly at Gamow energies~\cite{Bucher_2015PhRvL}, the lowest measured energy for the $\alpha$ and $p$ channels has been pushed only down to E$_{c.m.}$ = 2.1~MeV with large uncertainties.
Three different kinds of extrapolations have been used to estimate the reaction cross section at lower energies~\cite{cf88,beck_2020EPJA,Tumino_2018Nature,Jiang2018_hindrance,Fruet_2020,Zhang_2020PhLB,LI_2020_CCFUSION}.
A recent $^{24}$Mg($\alpha$,$\alpha'$)$^{24}$Mg experiment at RCNP provides important information for the resonances at stellar energies~\cite{Kawabata_2012}. Extrapolating theory is being developed at Malaysia and Japan.
The IMP group is carrying out a direct measurement using the intense carbon beam from the Low Energy Accelerator Facility(LEAF)~\cite{Yang_LEAF_2019PhRvS} while the LUNA collaboration is planning an underground measurement~\cite{Guglielmetti_2014_LUNA_MV}. High intensity heavy ion beam, efficient detectors with ultra-low background, theoretical studies of the molecular resonances and fusion reaction theory are needed to achieve a more reliable reaction rate for the astrophysical application.


\subsubsection{Explosive Hydrogen burning in Novae, X-ray bursts and Type II Supernovae}
Novae and type I X-ray bursts (XRBs) are the most frequently observed explosive stellar events. The observed light curves and isotopic/elemental abundances are important probes to study the compact stellar objects and the related nucleosynthesis. Due to a large systematic uncertainty of the theoretical results from estimated nuclear physics inputs such as nuclear mass and reaction rate\cite{Cyburt2016}, experimental knowledge of the constituent reactions that power these violent outbursts are required to improve the uncertainty.

Explosive nucleosynthesis involving the HCNO cycle and the rp-process (see Fig. \ref{fig:nucl_chart}) is thought to be responsible for the production of terrestrial $^{15}$N and the excess $^{22}$Ne seen in many meteorites, as well as the elemental over-abundances of O, Ne, Mg, etc. observed in nova ejecta. In order to understand the dynamics of such explosions and the origin of our solar system materials, it is necessary to determine the detailed characteristics
of the important nuclear reactions in the explosive process. Important ($\alpha$,p) reactions on $^{14}$O, $^{18}$Ne, $^{22}$Mg, and $^{30}$S, $^{15}$O($\alpha$,$\gamma$)$^{19}$Ne and (p,$\gamma$) reactions on $^{13}$N, $^{18}$F and $^{19}$Ne were studied with radioactive ion beams at CRIB, RIKEN-RIPS, HIRFL-RIBLL1 and HRIBF~\cite{Kubono_2002,Kim2015,Zhang_2014PhRvC_18neap,Kubono_1989,Motobayashi_1991,Kahl:2018}. However, most results are yet limited, and thus further studies are need.

The rp-process is a dominant mechanism of nucleosynthesis in XRB which is ignited in low mass binary systems consisting of a neutron star and a H/He rich companion star (See Fig.\ref{fig:nucl_chart}). When the critical temperature and density are reached during XRB, the rp-process is triggered by the 3$\alpha$-reaction and subsequently powered by a sequence of (p,$\gamma$) and ($\alpha$,p) reactions. This process burns materials in the star into the calcium region. After the process, proton capture and $\beta^+$-decay occur and produce the nuclei along the proton drip line up to A=100. It has been long considered that $^{64}$Ge, $^{68}$Se, and $^{72}$Kr nuclides are the major
waiting-point nuclei along the path of the rp-process. The predictive power of the XRB model is limited by the nuclear uncertainties around these waiting points which can only be removed with precise measurements. Taking the $^{64}$Ge as an example, it was pointed out that the uncertainties of the mass and the proton capture rate of $^{65}$As limited the predictive power of the XRB model. The mass uncertainty was eliminated by a precise mass measurement of $^{65}$As performed using the HIRFL-CSR\cite{Tu_2011PhRvL.106k2501T}. An experiment has been performed at RIKEN RIBF using the SAMURAI spectrometer, which measured $^{65}$As and proton decaying in flight from $^{66}$Se produced in one-neutron removal reaction induced by RI beams of  $^{67}$Se. The decay information was used to investigate their roles in the proton capture reaction on $^{65}$As which is involved in the breakout of the rp-process starting from one of the waiting point nucleus $^{64}$Ge in type-I XRBs.

The $\nu$p-process in type II supernovae might run a little closer to the line of stability because of presence of a small fraction of neutrons which discard the waiting points of the rp-process\cite{Wanajo_2011}. The neutrino oscillations, which is discussed in Sec~\ref{subsec:weak}, the critical nuclear reactions and nuclear properties are essential for the model development. A recent precision mass measurement at HIRFL indicated that the rp-process may go through the Zr-Nb region, suggesting less feasible to have a previously reported Zr-Nb cycle at $^{84}$Mo\cite{Xing_2018PhLB..781..358X}. $^{56}$Ni(n,p)$^{56}$Co has been identified as the critical reaction which impact the nucleosythesis\cite{Wanajo_2011}. This reaction will be studied indirectly by slowing-down beams with a novel technique in the OEDO (Optimized Energy Degrading Optics for RI beam) project, which is installed in RIKEN RIBF by CNS~\cite{OEDO_instrument}. 

Recently, astronomers reported observation of a bump near the K edge of nuclides of Z=48-49 in the XRB\cite{Kubota_2019}, although the resolution was not good. High-resolution X-ray observations are really awaited in the future. 

\subsubsection{Radioactive tracers of the galaxy}
\begin{figure}[htbp]
\centering
\includegraphics[width=0.9\linewidth]{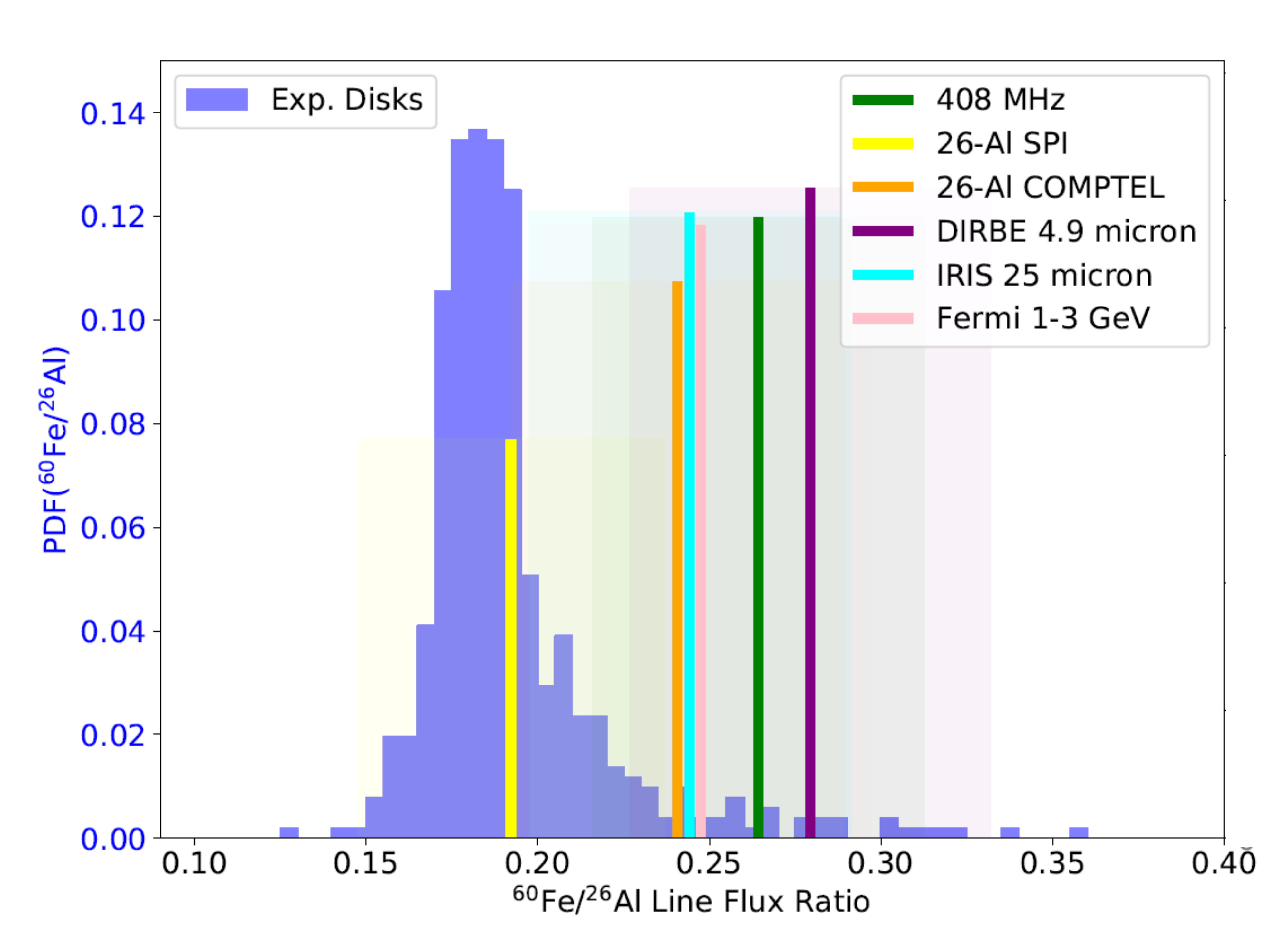}
\caption{\label{fig:fe60al26_ratio} $^{60}$Fe/$^{26}$Al flux ratio for the grid of exponential disk models (blue). Including
the uncertainties of the fluxes from each spectral fit, the total estimated $^{60}$Fe/$^{26}$Al flux ratio from
exponential disks is 0.184$\pm$0.042~\cite{wang_2020ApJ}.}
\end{figure}

The long-lived isotopes $^{26}$Al(T$_{1/2}$=0.72 Myr) and $^{60}$Fe(T$_{1/2}$=2.6 Myr) are produced mainly by massive stars. The nucleosynthesis of the long-lived radioactivities is an important constraint on the stellar models. The abundances inferred from gamma-ray astronomy may have important implications for rotationally induced mixing, convection theory, mass loss theory, the initial mass function for massive stars, and the distribution of metals in the galaxy. The present gamma-ray observations determine the $^{60}$Fe/$^{26}$Al gamma-ray flux ratio to be 0.184$\pm$0.042 based on the exponential disk grid maps~\cite{wang_2020ApJ}, as shown in Fig.\ref{fig:fe60al26_ratio}. But the mean ratio may vary in the range of 0.2 to 0.4 by using different sky maps. 
Theoretical groups predict the gamma-ray flux ratio at a large variation from 0.1 – 1\cite{Timmes_1995,Prantzos_2004,WOOSLEY_2007,Chieffi_2013,Sukhbold_2016}. As the major source of the large uncertainty comes from the reaction rates of the critical reactions, such as the neutron source reactions and those creation and destruction reactions for the unstable isotopes $^{26}$Al and $^{60}$Fe which have not be measured adequately, it is impossible to fully constrain the stellar model with the observed $^{60}$Fe/$^{26}$Al gamma-ray flux ratio.

Experimental and theoretical studies have been being carried out to identify the critical reactions and improve their accuracies with various techniques. Nucleosynthesis calculation identified the $^{59}$Fe stellar decay rates as one major uncertainty in the model\cite{lika_2016}. This uncertainty has been eliminated recently by an IMP--NSCL collaboration. Their experimental result shows the $^{60}$Fe yield would be reduced by using the new rate, alleviating the tension between some model predictions and the observation\cite{gao_2021}. The $^{59}$Fe(n,$\gamma$)$^{60}$Fe was re-studied by the CIAE and JAEA collaboration using the surrogate method. The $^{26}$Si(p,$\gamma$)$^{27}$P and $^{25}$Al(p,$\gamma$)$^{26}$Si reactions are well known as the relevant synthesis mechanism of $^{26}$Al$_{gs}$. The $^{26}$Si+p and $^{25}$Al+p elastic scattering experiments were performed by using CRIB to study the abundance of $^{26}$Al. Six resonance states were observed in the $^{26}$Si+p elastic scattering experiment. The reactions with the isomer state of $^{26}$Al in the stars complicates the nucleosynthesis of $^{26}$Al. Fortunately $^{26}$Al isomer beam becomes available at CRIB and the BRIF facility at CIAE for experiments. $^{25}$Mg(p,$\gamma$)$^{26}$Al is the important production reaction for $^{26}$Al. JUNA carried out a direct measurement to determinate the strengths and ground state feeding factors of the crucial resonances. 

Another long-lived isotope, $^{244}$Pu(T$_{1/2}$=80.0 Myr), are found in the deep sea sediment together with $^{60}$Fe using AMS\cite{Wallner_2015}. The current observation suggest that it comes from a rare event, such as neutron star merger. The consist understanding of the galactic radioactivity and deep sea sediment would help us understand the origin of the elements heavier than iron discussed latter in Ref.\ref{Topics:r}. It is also interesting to further study the moon samples from the Apollo or Chang-Er missions for finding the clues of the impact of SN events on the solar systems. 

\subsection{Neutron star and EoS of high density nuclear matter}
\begin{figure}
\includegraphics[width=0.9\linewidth]{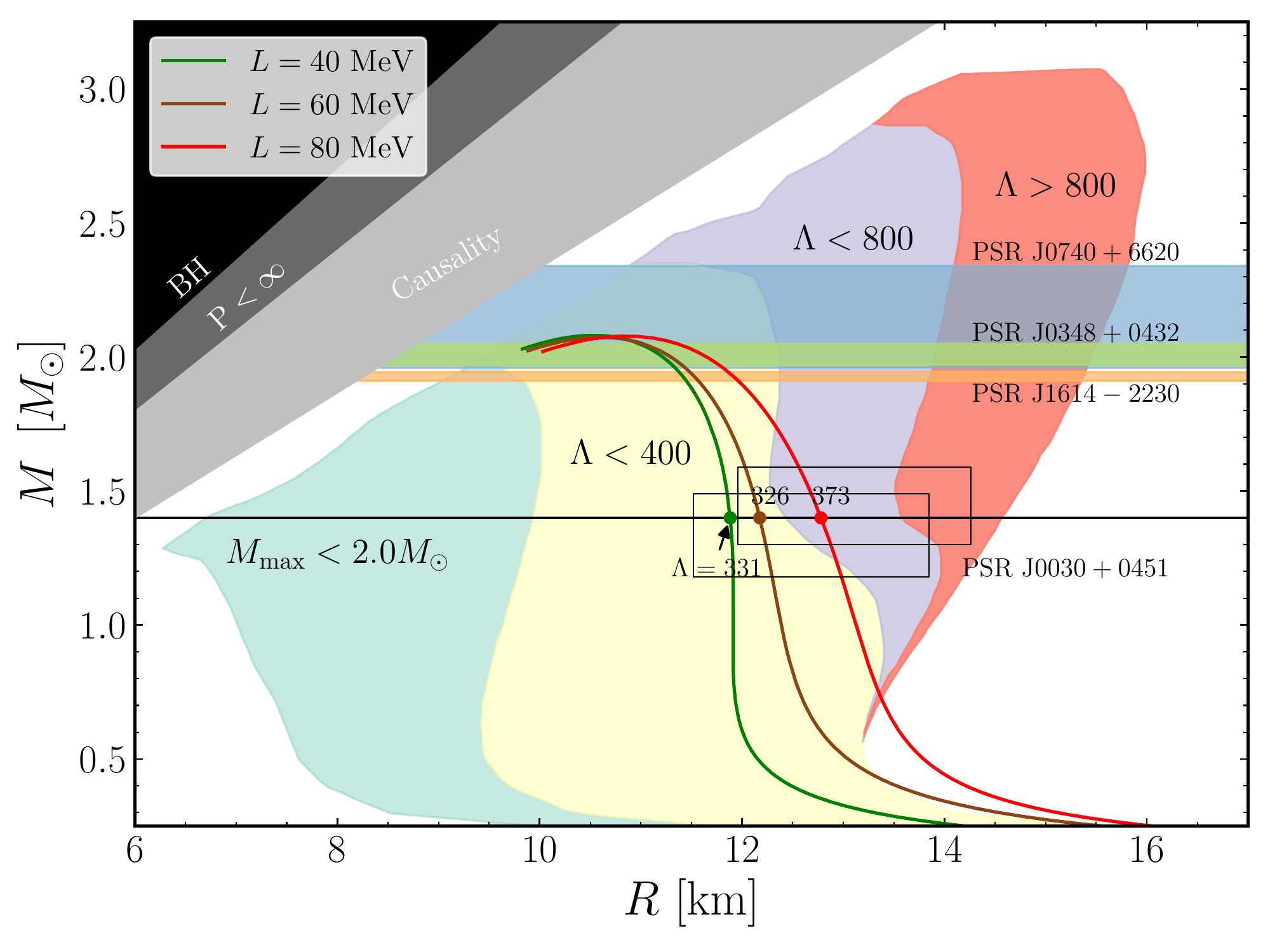}
\caption{NS mass-radius relation with different values of symmetry energy slope $L$ and the observational constrains from the  GW170718 NSM event and other NSs. Find detailed explanations in Ref.\cite{LI2020}.
}
\label{fig:mr}
\end{figure}

Neutron star(NS) is the smallest and densest stellar object composed of nuclear matter. It is kept in hydrostatic equilibrium only by the pressure produced by the compressed nuclear matter. Nuclear matter in neuron star varies from very low density on the surface to several times the saturation density in the core. The property of nuclear matter is often expressed in terms of the EoS. The symmetry energy term ($L$) of EoS is one of the key for the high-density nuclear matter. It determines the important NS properties such as pressure, structure, radius, tidal deformability ($\Lambda$) and so on (Fig.\ref{fig:mr}). The EoS of NS is also a fundamental component in the models of X-Ray Burst (XRB) and  the neutron star merger (NSM), a plausible site for the r-process nucleosynthesis. Thus understanding the NS EoS of dense matter is one of the fundamental quests of both nuclear physics and astrophysics.

Though the nuclear matter cannot be directly examined
in laboratories, to indirectly explore the nature of nuclear interactions 
in neutron-rich situations in finite nuclei is one of the target of experimental
nuclear physics. On-going efforts at RIKEN RIBF and
RCNP respectively aim at determining the symmetry energy term of EoS from 
different approaches: 
studies of multi-particle emission in collisions between neutron-rich nuclei
and evaluation of the neutron skin thickness through
giant resonance involving proton-neutron exchange. 
By colliding rare isotope Sn beams with isotopically enriched Sn targets, S$\pi$RIT collaboration deduced the slope of the symmetry energy to be 42 MeV$<$ $L$ $<$ 117 MeV~\cite{spirit_tpc_2021}, which is slightly lower but consistent with the L-values deduced from the  measurement of the neutron skin thickness of $^{208}$Pb.
Isolated multi-neutron systems and multi-neutron cluster states in neutron-halo nuclei are also investigated at RIBF. With the heavy ion beam covering sub GeV/u range, HIRFL-CEE
is being built at IMP to study EoS at about 2 times saturation
density via multi observable~\cite{Xiao_2014,CEE_2017SCPMA}. The inner core of
neutron stars is predicted to be composed of hyperon-mixed matter.
Its properties can be studied by the nature of hyper
nuclei at J-PARC. New opportunity will be available soon
at GSI and HIAF by studying the hypernuclear spectroscopy with
Heavy-Ion Beams~\cite{Saito_2016}. 

Lots of work have been done to connect the terrestrial experimental studies consistently with the observational results, like the mass constraint of the most massive pulsars, the tidal deformability constraints from GW170817 by LIGO/Virgo, and the simultaneous estimation of the mass and radius of PSR J0030+0451 by NICER, also incorporating our best knowledge of neutron matter based on ab initio calculations. Many efforts have been made on one of the EOS's main features, i.e., the symmetry energy, which is crucial for interpreting many astrophysical observations related to NSs and NS binaries. The future measurements of the radii of massive pulsars, using X-ray missions like NICER and eXTP, should provide even better constraints on the EOS~\cite{eXTP_2018}.


In the future, it is important to establish new quantitative results of the EoS that are testable by experiments/observations to understand the nuclear interaction better and to probe the particle degree of freedom in cold, dense matter.
It is also helpful to find further opportunities to study EoS from the dynamical processes of NS systems, like NS cooling, pulsar glitches, and short gamma-ray bursts.

\subsection{\label{Topics:r} r-process} 

Origin of radioactive nuclei such as $^{226}$Ra (half-life $\tau_{1/2}$=1600y), $^{232}$Th ($\tau_{1/2}$=140.5Gy), $^{238}$U ($\tau_{1/2}$=44.7Gy) is a long-standing unsolved question in modern science since the discovery of radioactivity by Antoine H. Becquerel in 1896 and the extraction of radioactive substances by Marie Curie in 1898.
These atomic nuclides are now thought to be produced in rapid neutron-capture process (r-process) in extremely neutron-rich environment of explosive phenomena~\cite{Fowler:1957}.
Binary neutron star mergers (NSMs) as well as core-collapse supernovae (CCSNe) and collapsars are the most viable candidates for the origin of r-process elements among many other possible astrophysical sites\cite{Kajino:2019,Cowan:2019pkx}. 
CCSN includes both neutrino-driven wind ($\nu$-wind) and magneto-hydrodynamic jet (MHDJ) SNe, which leave neutron star as a remnant, and collapsar is a very massive single star collapsing into a black hole~\cite{MacFadyen:1999,Siegel:2019}.

Three different aspects are folded in the studies of r-process nucleosynthesis. They are the nuclear properties of extremely neutron-rich short-lived radioactive nuclei, the theoretical modeling of SN explosion, dynamics of NSM and nucleosynthesis, and the spectroscopic observations of r-process elements in kilonovae caused by neutron star mergers as well as in metal-deficient stars to test these theoretical predictions. 

\subsubsection{Highlights from Experimental Nuclear Physics}

The r-process reaction flow path runs on extremely neutron-rich unstable nuclei 
far from the line of nuclear stability.  Recent innovative experimental technique in nuclear physics has reached the production of some of these exotic nuclei.
The fission of $^{238}$U is a possible powerful tool for producing very neutron-rich nuclei close to or on the r-process pathway. RIBF has been developing a high-energy and high-intensity $^{238}$U beam, and various secondary beams of unstable nuclei through in-flight fission allows one to study the r-process nuclei.  

\begin{figure}
\includegraphics
[width=1.0\linewidth]
{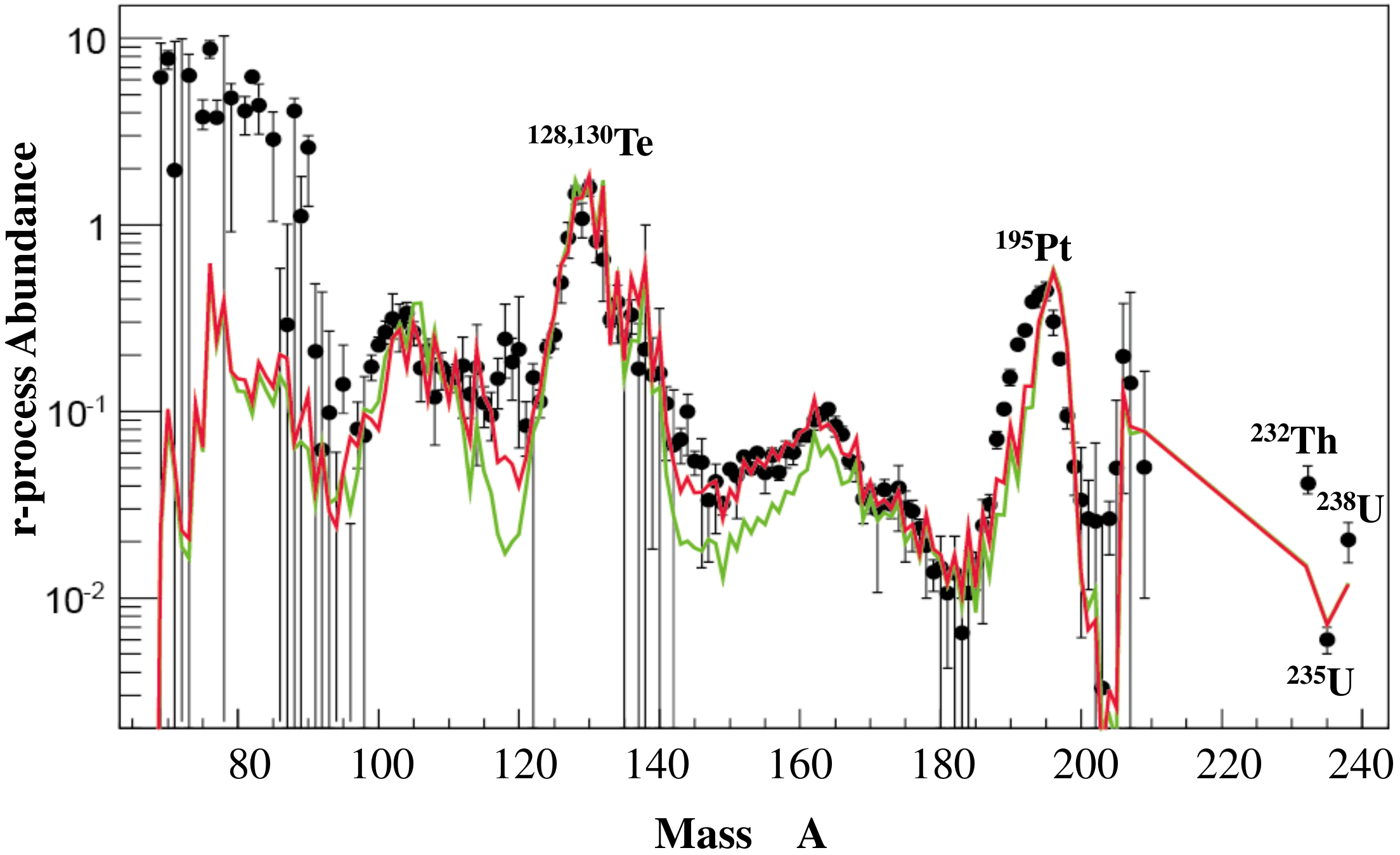}
\caption{\label{fig:abundance_RIBF} A comparison between the estimated r-process abundance curves with (red) and without (green) taking account of the new experimental data from RIBF~\cite{LorussoG:2015}. Theoretical abundances are calculated in a canonical SN nucleosynthesis model assuming (n,$\gamma$) - ($\gamma$,n) equilibrium~\cite{Kajino:2019}.}
\end{figure}

The large amounts of lifetime data for $\beta$ decay have been taken at RIKEN RIBF~\cite{LorussoG:2015,Wu:2017,Wu:2020}. They now cover the r-process pathway below and beyond the $N$=82 magic number.  The impact of the RIKEN data is demonstrated in Fig.~\ref{fig:abundance_RIBF} in a canonical SN nucleosynthesis model, where a network calculation assuming (n,$\gamma$) - ($\gamma$,n) equilibrium using RIKEN data available in 2015 is compared in two cases with and without the new experimental lifetimes~\cite{LorussoG:2015}. One can see the difference clearly. 
Data at around another waiting point $N$=50 also are available~\cite{XuZY:2014}.

In neutron-rich nuclei, the $\beta$ decay daughter can be in its high lying excited state that emits one or more neutrons, which influences the final abundances.
The BRIKEN project has started to measure the neutron emission probability $P_{\rm n}$, and a first result has already been obtained for a decay of $^{86}$Ge affecting the 
decay paths to the production of $^{84}$Kr, $^{85}$Rb and $^{86}$Kr. 

Nuclei with neutron magic numbers, which are called "waiting points", are expected to play an important role in forming the r-process abundance peaks. 
Spectroscopic studies at RIBF could access the "doubly-closed" $^{78}$Ni with $Z$=28 and $N$=50 for the first time. $^{78}$Ni is a typical seed nucleus for the r-process in some stellar conditions. Its low lying states have been identified by measuring deexcitation $\gamma$ rays from fast moving $^{78}$Ni nuclei produced by a few nucleon removal reaction in inverse kinematics, where RI beams hit a stable target~\cite{TaniuchiR:2019}. Together with lifetime measurements of nuclei in the vicinity of $^{78}$Ni~\cite{XuZY:2014}, the closed shell nature was found to persist with no clear signal for "shell quenching"~\cite{nishimura:2012,Kajino:2017}. 

Another neutron-rich region of interest is around $N$=82, responsible for the 2nd peak of the r-process abundance. 
Nature of 
palladium isotopes around the r-process path have 
been studied by measuring $\gamma$ rays from their excited states populated either by isomer production~\cite{WatanabeH:2013} or nucleon removal reaction~\cite{WangHe:2013}. Measured systematic trend of the low-lying state indicates no significant shell quenching for $N$=82.

Using the low-energy RI beams in the OEDO project, (d,p) reactions as surrogate process of corresponding (n,$\gamma$) reactions on r-process nuclei will be studied under the a collaboration involving Asian institutes such as CNS, RIKEN, IMP, and SKKU.  The region around $^{132}$Sn with $N$=82 is a target.
The feasibility of the method was already tested successfully for  $^{79}$Se(n,$\gamma$) at around few MeV of neutron energy.  

The KISS project and its future extension KISS-II aims to approach the blank-spot or the "south-east" region of $^{208}$Pb and further "east" beyond $N$=126, which includes the progenitor parent nuclei for the r-process 3rd peak element.  The group of Wako Nuclear Science Center (WNSC) of KEK found the deep-inelastic collisions or multi-nucleon transfers can be more efficient to reach that region compared with the projectile fragmentation or fission usually employed in most of the RI-beam facilities.  The project has started by building a system~\cite{HirayamaY:2017}.

As illustrated in Fig.~\ref{fig:nucl_chart}, fission fragment distribution (FFD) as well as various fission modes, i.e. spontaneous, $\beta$-delayed and neutron-capture induced fissions, are known to affect strongly the r-process abundance 
in NSM and collapsar nucleosynthesis~\cite{Shibagaki:2016,suzuki:2018,famiano:2020}. It is planned to systematically measure the fission barrier of the neutron-rich actinoids 
with the SAMURAI setup at RIKEN RIBF. Nuclear data groups at Japan Atomic Energy Agency and Tokyo Institute of Technology successfully describe the FFD and their kinematical properties for long-lived radioactive actinoids and SHEs in a unique method of solving 3D/4D Langevin equation~\cite{HiroseNishio:2017,UsangChiba:2019}.  

Measurement of nuclear masses is essential to study the nuclear structure date relevant for r-process~\cite{ZhaoArima:2012,Sun:2008}.
In addition to the MR-TOF method in the SLOWRI setup for slow RI ions and the attempts to measure TOF in a long flight path, a novel method is being developed.
Individually injecting short-lived nuclei are produced and identified in BigRIPS and their masses are determined by realizing the isochronous condition of the Rare RI Ring. 
The goal is to achieve 10$^{-6}$ mass precision for nuclei with a lifetime as short as 1ms, which should greatly improve our knowledge on the r-process network.

\subsubsection{Highlights from Astronomical Observations and Theoretical Progress}

Follow-up observations of optical-to-infrared light from GW170817, including the contributions of the observations with the Subaru Telescope lead by the J-GEM consortium~\cite{2017PASJ...69..101U}, have provided an evidence that some sort of radioactive nuclei have been produced in binary NSM. The infrared excess over the optical spectra at several days was interpreted as the signature of large opacity due to the lanthanoids. 
Theoretical modeling of NSM r-process suggests that both the lighter and heavier r-process elements could be produced in the early dynamical ejecta and/or the later wind outflow from the remnant~\cite{Wanajo:2014,Thielemann:2017,Cowan:2019pkx}. However, recent careful re-analysis of the spectra~\cite{Watson:2019} and many other related studies (e.g., \cite{Kasen:2017sxr,2020MNRAS.496.1369T}) have indicated that the amount of lanthanoids is not as large as that expected in theoretical calculation for kilonova associated with GW170817. 
Meanwhile, the detection of the infrared emission at several tens of days has not yet been able to provide evidence for the presence of $^{254}$Cf or $\alpha$-decay nuclei. However, such association can be possible with future events~\cite{Wu:2019xrq,Zhu:2018oay}.

Although theoretical model uncertainties still remain, there are large parameter space to make variations of the NSM r-process as a possible astrophysical site of the r-process~\cite{Cowan:2019pkx,Shibagaki:2016,Thielemann:2017}. 
The basic difference between NSM and CCSN is a different emergent event rate and a delay timescale from star formation of their progenitors until the time when r-process occurs. SNe and collapsars are dying single massive stars that culminate their evolution in explosions in a few My. Therefore, they can enrich r-process elements in the interstellar medium (ISM) from the early Galaxy. 
On the other hand, the majority of NSMs may not contribute to the early Galaxy because they can take cosmologically long timescale due to extremely slow GW radiation to lose kinetic energy of the orbital motion until they merge.  The coalescence timescale estimated from observed double pulsars ranges from a few hundred My to the cosmic age or beyond~\cite{Swiggum:2015} although faster mergers are also possible. Their emergent rate is as low as 0.1-1\% of SN rate. Recent theoretical study of Galactic chemical evolution has demonstrated that it may be difficult for NSMs to emerge in metal-deficient region for these reasons~\cite{Yamazaki:2020}. 

The contribution from NSMs to r-process elements, however, depends on the timescale of metal enrichment and dynamics in chemical evolution models. Recent models include formation of metal-poor stars in small stellar systems like the current dwarf galaxies around the Milky Way, which takes longer timescale for chemical enrichment in general until accreting into the Galactic halo. Such models at least partially reconcile the difficulty of the timescale problem of NSM contribution~\cite{Hirai2015}. Recent discoveries of r-process-enhanced stars in dwarf galaxies like Ret.~II are providing useful constraints on such models (see below).

Observations of early generations of stars also made progress in revealing the nature of the r-process and identifying the sites. 
Chemical compositions of heavy neutron-capture elements show large star-to-star scatter, among which a small fraction of extremely metal-poor (EMP) stars have large excess in r-process elements like Eu~\cite{Ishimaru:1999}. This indicates that these heavy elements are produced by quite rare events and that the gas clouds were not well mixed in the early universe~\cite{Sneden:2008}. Ultra-faint dwarf galaxy named Ret.~II which has just thousands of solar masses was discovered, and interestingly most stars seem to be r-process-enhanced~\cite{Ji:2016,Roederer:2016}. This suggests that the progenitor gas cloud was polluted by a few but very efficient r-process events, which could be NSMs, MHDJs or collapsars. It is suggested in a hierarchical structure formation scenario that the Milky Way halo was formed from merging dwarf galaxies which consists of r-process enhanced stars. 

China-Japan collaboration team has recently reported the discovery of an r-process-enhanced star with relatively high metallicity compared to those found previously \cite{2019NatAs...3..631X}. The metallicity of this object is about 30 times lower than the Sun, but its Eu abundance is as high as the solar value. Whereas such object had not been found previously in the Milky Way, similar objects are known in satellite dwarf galaxies \cite{2007PASJ...59L..15A}. This discovery suggests that the  hierarchical structure formation scenario of the Milky Way is applicable to relatively metal-rich halo stars, as mentioned above, such that the objects were formed in dwarf galaxies first and then accreted later into the Milky Way. Interestingly, objects having similarly large excess of r-process elements have been found among moderately metal-poor stars ($-2<$[Fe/H]$<-1$) in the past two years. 

The detailed abundance patterns determined for individual objects provide strong constraint on understanding of r-process nature. A remarkable result found for r-process-enhanced EMP stars is that their abundance patterns of heavy elements are very similar to each other, and also similar to that of the solar-system r-process component, which is called "universality" of the r-process~\cite{Sneden:2008}. 
This empirical fact invited many theoretical models of SN r-process, e.g.~\cite{Wanajo:2014,Otsuki:2000,Terasawa:2001}, as shown in Fig.~\ref{fig:abundance_RIBF}. Recent calculations have suggested that isotopic abundance patterns as a function of nuclear mass number A are quite different although elemental distributions as a function of atomic number Z are similar to one another among CCSNe, collapsars and NSMs~\cite{Shibagaki:2016,Yamazaki:2020}. Isotopic separation of r-process elements in EMP halo stars are highly desirable.
On the other hand, diversity of abundance patterns of light neutron-capture elements also are found in metal-poor stars~\cite{2017ApJ...837....8A}. Further theoretical studies of explosive nucleosynthesis and observations of SN and GW events are highly desirable to uncover the origins of neutron-capture elements and the mechanisms to produce diversity in their abundance ratios. 

\subsection{\label{subsec:weak}Nuclear electroweak response and $\nu$-astrophysics}

Nuclear electroweak processes affect various masses of stellar evolution from the main sequence stage until the white dwarfs or gravitational collapse of massive stars~\cite{Suzuki:2016,Mori:2020blue,Mori:2020obx}. Neutrinos play an essential role in successful explosion of CCSNe by heating the gas behind the shock~\cite{OConnor:2018sti} and affect various explosive nucleosyntheses. Since solar neutrinos are the direct observable to exhibit nuclear fusion inside invisible solar interior, their observed flux serves to construct the standard solar model~\cite{Christensen-Dalsgaard:2020imv}. Nuclear electroweak response has been studied in laboratory experiments by measuring nuclear $\beta$-decays and hadronic charge-exchange (CEX) and photo-nuclear reaction cross sections because neutrinos are thus invaluable messengers of weak processes in celestial events. The CEX and photo-induced reactions are analogous to charged and neutral current neutrino-induced reactions, respectively.

Systematic theoretical calculations of the neutrino-nucleus reaction cross sections have been performed based on sophisticated theoretical models such as nuclear shell models using new generations of Hamiltonian based on the progress in physics of exotic nuclei~\cite{Suzuki:2018aey,TanSun:2020} and quasi-particle random phase approximation (QRPA)~\cite{Cheoun:2010,Cheoun:2011hj}. The results very well explain the experimental data on $^{12}$C~\cite{Athanassopoulos:1997rm} and $^{56}$Fe~\cite{Bergeet91,Oltman92}, and these models have been extensively applied to provide unmeasured $\nu$-nucleus reaction cross sections of astrophysical interest.

Neutrino experiments in Asian countries/regions have contributed remarkably to determine the mixing angles, i.e., $\theta_{12}$ (Super-KAMIOKANDE, KamLAND), $\theta_{23}$ (Super-KAMIOKANDE, K2K, T2K) and $\theta_{13}$ (T2K, Daya Bay, RENO) although mass hierarchy and CP-violation phase are still unknown. Upon these experimental efforts, it is the recent focus to determine the unknown oscillation parameters in terms of SN expolosions that provide three-flavor neutrinos and their anti-particles~\cite{2015PhRvD..91f5016W}. 

The neutrinos emitted from proto-neutron stars propagate through the stellar interior where several nuclei are produced by $\nu$-nucleus interactions. The nuclides heavier than iron are produced in the $r$- and $\nu p$-processes in neutrino-driven winds\cite{Pruet:2004vb,Frohlich:2005ys,Wanajo_2006,Sasaki:2017jry} and the $\nu$-process in outer layers~\cite{Domogatskii78,Woosley:1989bd,Heger:2003mm,Yoshida:2006qz}. Neutrinos also play an important role in the NSM outflows where a wide mass rage of $r$-process nuclei are produced\cite{Wanajo:2014,Just:2014fka,2016MNRAS.463.2323W} 
The $^7$Li and $^{11}$B are particularly important~\cite{Yoshida:2006qz}. A recent improved and more consistent calculation of stellar evolution from He star to SN explosion has found a dependence of their
yields on stellar metallicity~\cite{Kusakabe:2019znq}.

\begin{figure}[ht!]
    \begin{center}
      \includegraphics[width=7.0cm,clip]{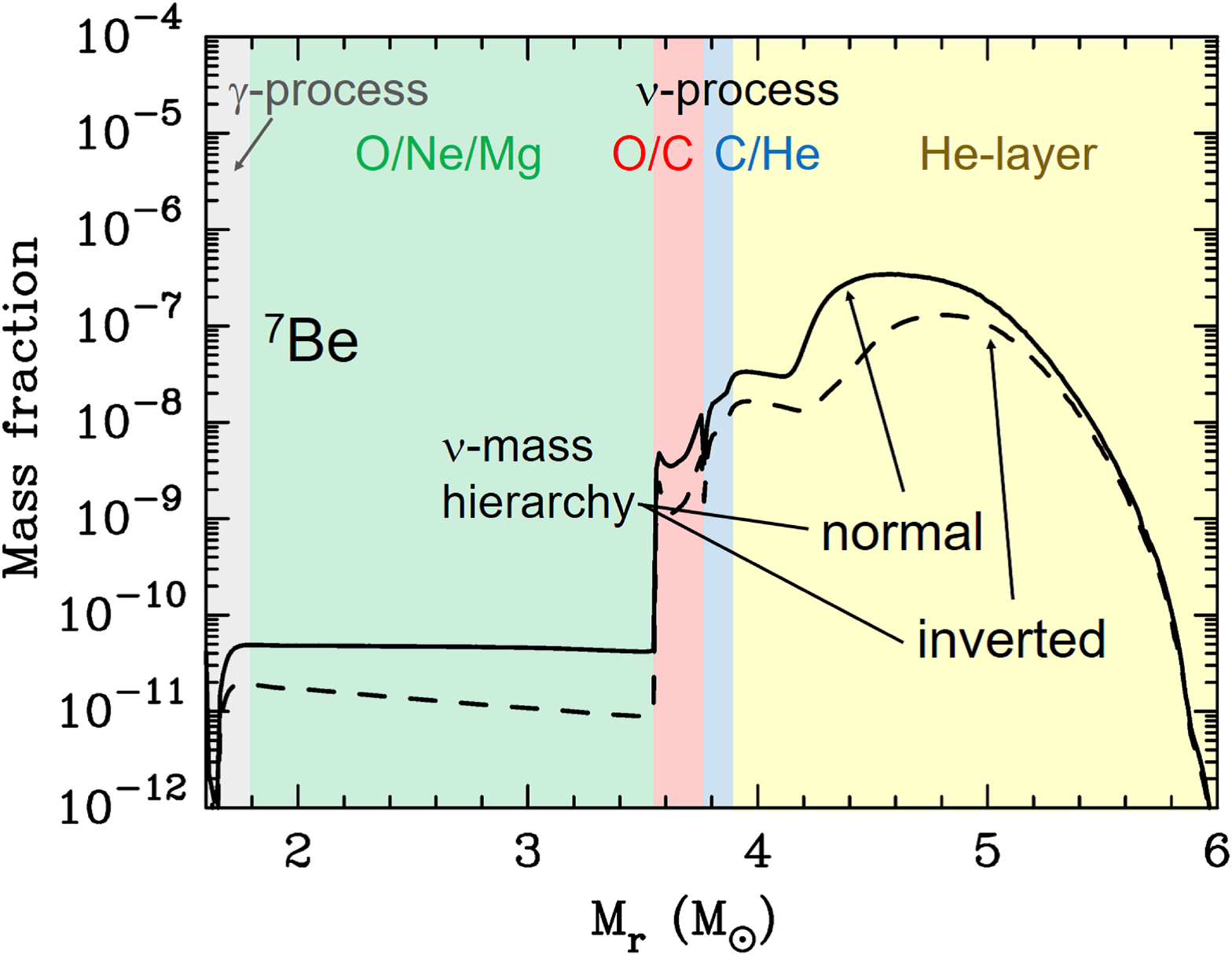}
      \includegraphics[width=7.0cm,clip]{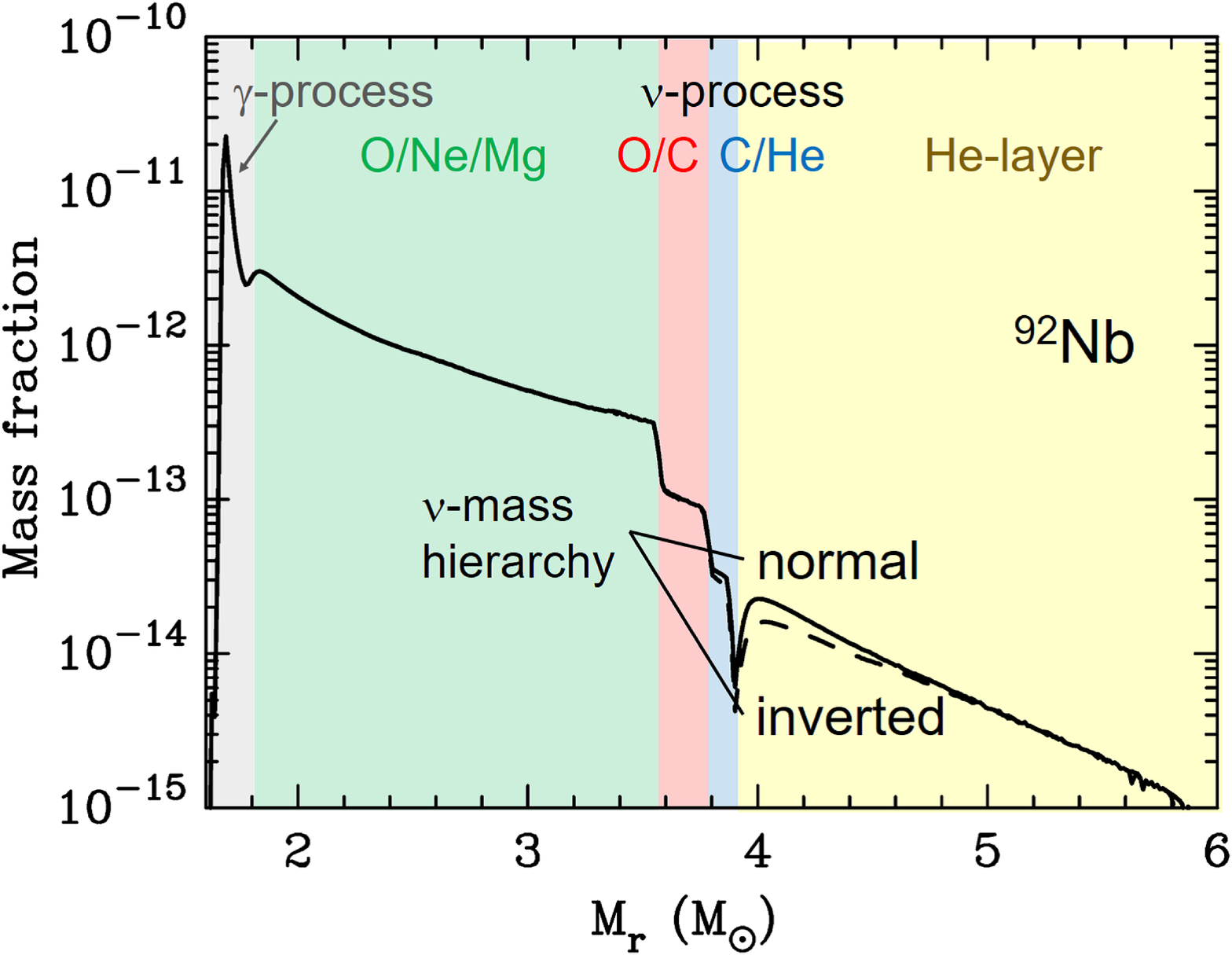}
      \caption{Mass fractions of $^7$Be (left panel) and $^{92}$Nb (right) versus $M_r$ in SN 1987A model star. Solid and dashed curves are the results for normal and inverted mass hierarchies, respectively~\cite{Kusakabe:2019znq}. Colors distinguish $\gamma$-process region at high temperature, O/Ne/Mg, O/C, C/He, and He-rich layers for $\nu$-process region as labeled. $^7$Be is mainly produced in the outer He-layer, while $^{92}$Nb is produced also in the central hot region via the $\gamma$-process as well as $\nu$-process.}
      \label{fig1}
    \end{center}
  \end{figure}

It was found for the first time that remarkable amount of radioactive isotope $^{98}$Tc is produced via charged-current reactions by $\bar{\nu}_e$ although the other nuclei are produced mainly by the $\nu_e$-nucleus interactions. Thus, $^{98}$Tc is characterized as a special probe of SN $\bar{\nu}_e$ spectrum~\cite{Hayakawa:2018ekx}.
Extensive calculations showed that another short-lived radioactive isotope $^{92}$Nb and stable ones $^{138}$La and $^{180}$Ta also are produced in the $\nu$-process. These nuclei are useful to constrain still unknown mass hierarchy~\cite{Hayakawa:2010,Ko:2020rjq,Ko:2019asm}.
Fig.~\ref{fig1} shows an example of the $^7$Be and $^{92}$Nb abundances as a function of the Lagrangian mass $M_r$. The high-density MSW effect~\cite{Wolfenstein:1978,MikheyevSmirnov:1985} occurs near the bottom of the He-layer. On the other hand, the collective flavor oscillation induced by coherent $\nu$-$\nu$ scattering\cite{Duan:2010bg,Pehlivan:2011} occurs inside the iron core. Therefore, $^7$Be which is mainly produced in the He-layer is affected by both MSW and collective oscillation effects, while $^{92}$Nb which is predominantly produced in the O/Ne/Mg-layer is affected by only the collective oscillation effect. Neutrino mass hierarchy is thus imprinted in each isotopic abundance in a different specific manner. Almost all these nuclei are ejected in SN explosions, from which the silicon-carbide pre-solar grains called SN grains condensate. Meteoritic measurement of SN grains is highly desirable to find a piece of evidence for the neutrino mass hierarchy~\cite{Ko:2019asm,Mathews:2012}.  

Neutrino also holds the keys in the early universe because the cosmic expansion rate depends on the generations of light neutrino family~\cite{Shvartsman1969}. Several short-baseline neutrino oscillation experiments reported an anomaly in the measured flux which could be explained by fission-induced reactions or assuming the non-active fourth-generation neutrino which is called ''sterile neutrino''~\cite{Boyarsky:2018tvu}. Reactor neutrino experiments at Daya Bay (China), RENO (South Korea) and JSNS$^{2}$ (Japan) are going to look for the signal of sterile neutrinos in sub keV region. 

The dark energy and dark matter are the other key ingredients of the universe. Dark matter candidates include exotic weakly interacting massive particles (WIMPs) such as SUSY particles, right-handed massive neutrinos and axions. These hypothetical particles could be directly detected via the elastic scattering off the nuclei or atomic electrons through the weak interaction. Several experiments for WIMP hunting are underway at CDEX (China), PandaX (China), COSINE (South Korea), XMASS (Japan), NEWAGE (Japan), and PICO-LON (Japan).

\section{\label{sec:future}Future Facilities}
New facilities such as large telescope, neutrino observatory, next generation of Radioactive Ion Beam facilities and underground nuclear astrophysics experiments are being built or planned. More exciting discoveries are awaiting for us in coming decades. Here we briefly report some progresses of the facilities and related scientific problems.




\subsection{Multi-messenger Astronomy}
\label{subsec:multi}

The first successful detection of the neutrinos from SN1987A with KAMIOKANDE-III, IMB and Baksan experiments opened the door to neutrino astronomy, which was followed by successive discovery of neutrino oscillation. This event of the century enriched our understanding of explosive nucleosynthesis and the roles of neutrino interactions in SNe. Supernovae provide heat and matter into space and drive the chemical and dynamical evolution of galaxies in the cycle of star birth - stellar evolution - explosion - next generation of star birth~\cite{Nomoto:2013}. One of the current targets in neutrino astronomy at the underground laboratories is to detect relic SN neutrinos~\cite{Hidaka18}, which hints to determine the SN event rate in the Milky Way. In Japan, Super-KAMIOKANDE detector is in operation with improved sensitivity by adding gadolinium into the Cherenkov medium made of light water. Next project Hyper-KAMIOKANDE with remarkably upgraded size of water Cherenkov detector has started. In China, Jiangmen Underground Neutrino Observatory (JUNO) is under construction.  
 
\begin{figure}
\includegraphics
[width=0.8\linewidth]
{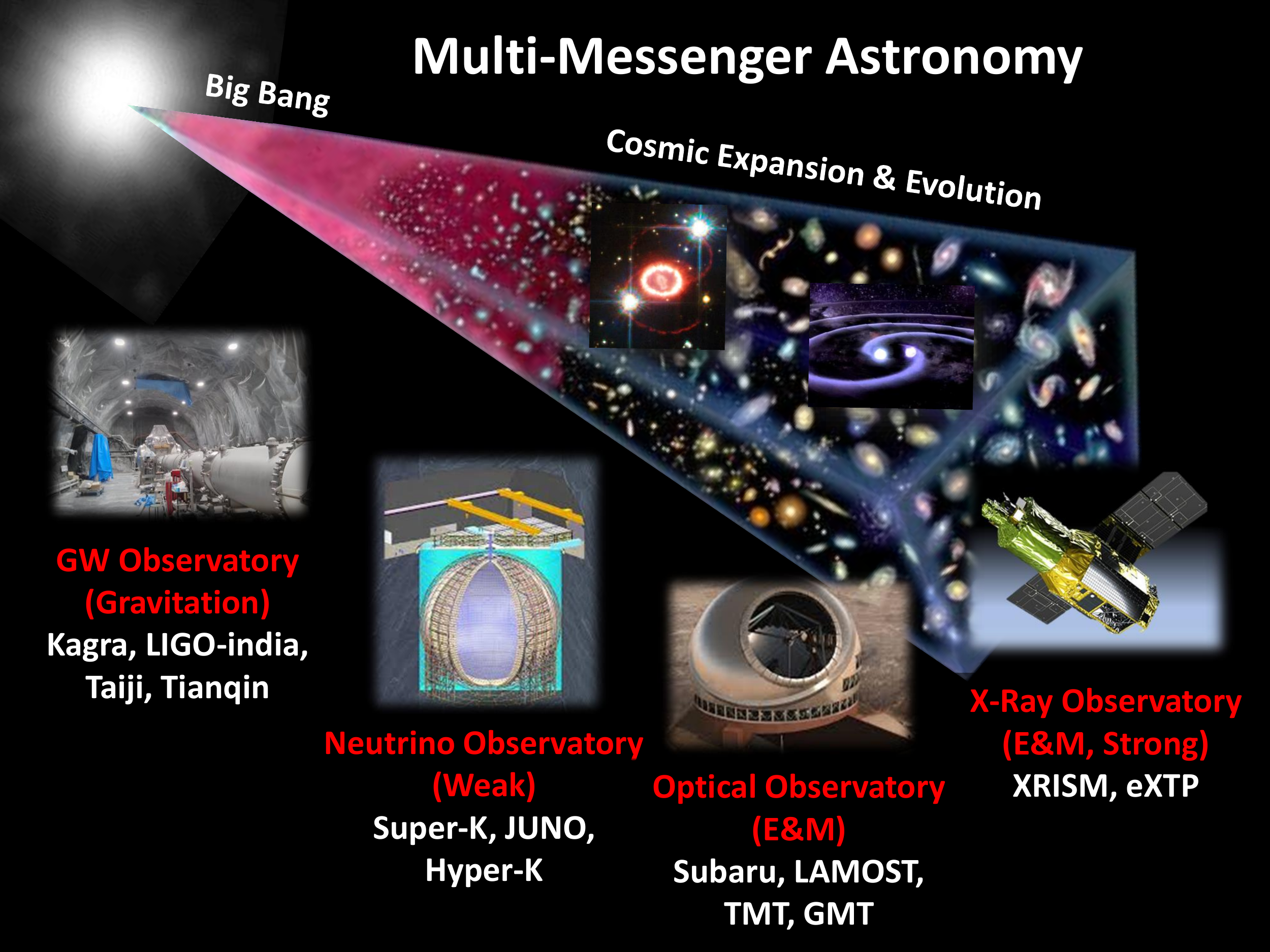}
\caption{\label{fig:multi-m} Planned future observatories under the ground, on the ground and in the space to detect multi-messengers from the cosmos. GWs, neutrinos, optical-to-infrared photons, and X$\gamma$-rays are respectively the messengers of gravitational, weak, electromagnetic, and electromagnetic plus strong nuclear forces that make a variety of physical and chemical processes in the universe, galaxies and stars.}
\end{figure}

GW astronomy will link to neutrino astronomy. GW170817 made a big impact on the studies of r-process nucleosynthesis and EoS of neutron stars. Simultaneous detection of both GW and neutrinos in the future would provide more details how the elementary particle and nuclear processes go on in catastrophic SN explosions as well as merging phenomena of relativistic compact objects. KAGRA in Japan has started operation~\footnote{https://gwcenter.icrr.u-tokyo.ac.jp/en/}, LIGO-India currently under construction~\cite{Adya:2020}, and the space-based GW detection projects TianQin~\cite{TianQin:2016} and Taiji~\cite{Taiji:2017} are planned in China.

Optical-to-infrared spectroscopic observations of SN1987A and GW170817 also gave fascinating insights into nucleosynthesis of heavy nuclei and dynamics ejecta. SN1987A appeared very close to us in Large Magellanic Cloud at a distance of 140kpc, and both multi-wave length photons and neutrinos were successfully detected. Although GW170817 appeared at 40Mpc which is relatively close and bright enough for GW detection and spectroscopic observations, it was too far to observe neutrinos. Future follow-up spectroscopic observations require deeper observations. High-resolution spectroscopic observation of faint stars in the Milky Way is planned with Subaru Telescope linking to large optical survey with LAMOST. Construction of next generation extremely large ground-based telescopes like Thirty Meter Telescope (TMT), Giant Magellan Telescope (GMT) and the space telescope like JWST are planned as international collaboration projects including Japan, Korea, China, India and Australia. 

The last piece of multi-messengers is the messenger of strong nuclear force and electromagnetism in the atomic and nuclear processes in the cosmos. Space X-Ray satellite mission XRISM (X-Ray Imaging and Spectroscopy Mission) is planned at ISAS in Japan as post HITOMI project after the successful X-ray satellite missions of HAKUCHO, TENMA, GINGA, ASKA, and SUZAKU in the past. The enhanced X-ray Timing and Polarimetry mission (eXTP) is planned to be launched in 2027 to study the state of matter under extreme conditions of density, gravity and magnetism by the collaboration between Chinese and European institutions. Technology in analyzing chemical compositions of meteorites or returned sample from asteroids like Ryugu in Hayabusa-II project~\footnote{https://www.hayabusa2.jaxa.jp/en/} would bring new insight how our solar-system formed and what the primordial solar material was made of. These might bring a piece of evidence for solving the mystery why optical chirality of amino acids on Earth is broken to be only left-handed~\cite{Famiano:2018}, which hints the creation of life. However, as for the origin of elements in the cosmos, these are still limited to nearby solar system. Complementary spectroscopic observations of various ages of stars await to seek for the cosmic and galactic evolution of atomic nuclides from the big-bang era until the solar system formation. 

It relies highly on the theoretical progress, too, in modeling the big-bang universe and high-energy phenomena like SNe and NSMs including numerical simulations of explosive nucleosynthesis there, as illustrated in Fig.~\ref{fig:multi-m}. New-generation supercomputers such as Japanese Fugaku and Chinese Tianhe and Sunway Taihu-Light have enormously big memory and extremely high-speed computing power so that they can simulate cosmic, galactic and stellar evolution and associated nucleosynthetic phenomena by integrating all observed information and knowledge to be brought into multi-messenger astronomy and astrophysics. 

\subsection{Future Facilities in Experimental Nuclear Physics}
We have several world-class facilities for radioactive ion beams existing in east and southeast Asia to study important nuclear astrophysics problems. Moreover, there are several upgrades and constructions toward more powerful facilities. Among them, upgrades of RIKEN RI Beam Factory (RIBF) and J-PARC in Japan, Rare Isotope Accelerator complex for ON-line experiments (RAON) in Korea, and High-Intensity heavy ion Accelerator Facility (HIAF) in China are introduced in this section.

RIBF facility has been in operation for more than ten years with its world's best capability of producing rare isotopes far from the stability. A variety of experimental studies related to nuclear astrophysics have been performed with RI beams of them (or these isotopes), as briefly described in previous sections. The upgrade of RIBF is being planned to continuously provide the best beam quality to our nuclear physics community in the world, together with upcoming facilities, such as FRIB in USA, FAIR in Germany, RAON and HIAF. 
A new charge-stripping scheme with specially-designed ring devices will transmit beam ions with multiple charge states, and recover the decrease of the beam in the two existing strippers. As a result, one order of magnitude higher beam intensity is expected~\cite{Kamigaito2020}.
Installation of a new superconducting LINAC, other improvements of the accelerators and the fragment separator BigRIPS will increase the accessibility of more neutron/proton-rich nuclei including isotopes related to the r-process.

J-PARC is a proton accelerator facility, and the intensive high-energy proton beams are used to produce high-flux secondary beams of neutrons, neutrinos and mesons such as kaons and pions. Astrophysics oriented investigations such as (n,$\gamma$) type reactions involved in the s-process are studied as well as properties of hyper-nuclei, hyperon-nucleon scattering. Neutrinos produced in the pion beam are also used to study neutrino oscillation with the KAMIOKA neutrino detector. The current extension plan for the hadron beamline will lead to systematic studies of strangeness nuclear physics and hence shed light on the hyperon-mixed matter.

\begin{figure}
\includegraphics
[width=0.95\linewidth]
{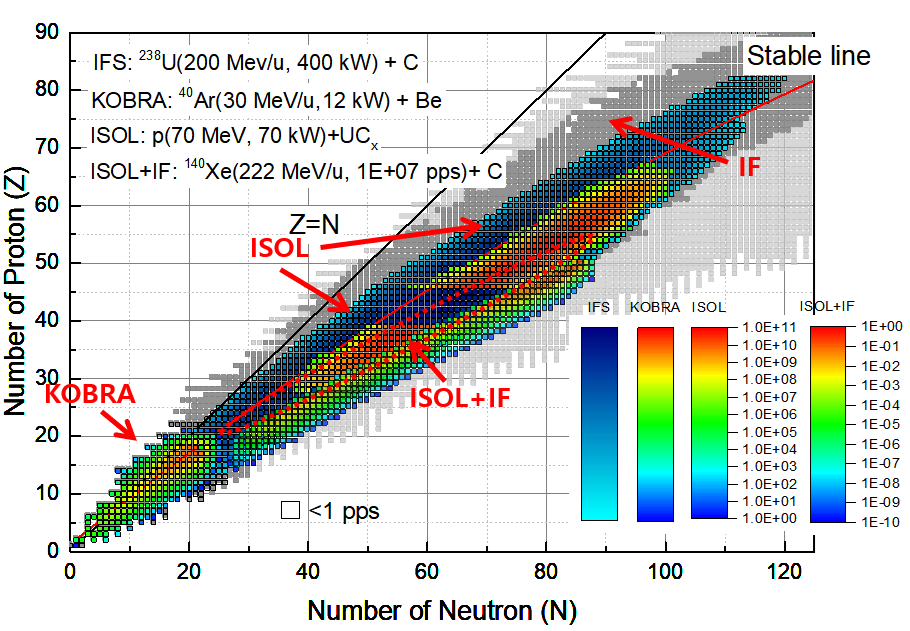}
\caption{\label{fig:RAON_yield}(Color online) The nuclear chart with expected rare isotope beams and their rates in unit of pps, produced by the KoBRA spectrometer, ISOL, IF and ISOL+IF methods at RAON.}
\end{figure}

One of the main scientific objectives of the RAON accelerator facility is to perform nuclear astrophysics experiments using RI beams. While RAON is designed to produce RI beams using known techniques including spectrometer, Isotope Separation On-Line (ISOL) and In-flight Fragmentation (IF), the most remarkable configuration is to utilize the combination of ISOL and IF system to produce more exotic species of isotopes. The method is very unique to provide neutron-rich beams with a greater intensity than that achieved by any existing facilities. Expected RI beams and their intensities are shown in Fig.\ref{fig:RAON_yield}. Details of the beam production methods are described in Ref.~\cite{Jeong2016}.

The KoBRA at RAON is a multi-purpose mass separator for studying nuclear structures and reactions of exotic nuclei in the energy range of less than 30 MeV/u, as well as nuclear astrophysics experiments. The KoBRA project is divided into two stages. The first stage of KoBRA, will be utilized to produce RI beams using stable beams for the early phase of RAON. KoBRA will be extended by having more beam line components and a spectrometer, which is the second stage, for the study of astrophysically interesting reactions including $^{15}$O($\alpha$,$\gamma$)$^{19}$Ne and $^{14}$O($\alpha$,p)$^{17}$F reactions.

HIAF is a new accelerator facility in China, aiming to provide high intensity heavy ion beams from a LINAC with a maximum current up to 1 emA. With a booster, the beam energies can reach 9.3 GeV/u for proton and 0.8 GeV/u for $^{238}$U. The lower energy terminal after the LINAC will be mainly used to produce new isotopes/elements using multiple nucleon transfer reactions or fusion reactions for the studies of nuclear reaction and spectroscopy. The research program with the higher energy beam from the booster will focus on the production of neutron-rich isotopes, precise mass measurements of exotic nuclei, charge exchange reactions, properties of hypernucleus, EoS of nuclear matter and the QCD phase diagram. The phase-I construction will be completed by 2025. China Initiated Accelerator Driven Subcritical System (CIADS) is also being constructed adjacent to HIAF to transmute nuclear waste using a powerful proton beam from the LINAC. In phase-II, an ISOL terminal will be built at CIADS to feed the neutron rich isotopes into HIAF. More critical r-process nuclei will be produced by the projectile fragmentation reaction or multiple nucleon transfer reaction induced by the neutron-rich beams.

Furthermore, successful collaborations among different facilities are crucial since the teamwork will lead the project development in a very efficient way and improve the performance of the experimental instruments.
New experimental techniques such as silicon detector arrays, $\gamma$-ray detector arrays and active target time projection chambers are critical for the success of the facilities. Collaborative efforts among the facilities will not only be very helpful to solve many  challenging problems during the developments of novel technologies in each institute, but also offer new possibilities to operate expensive arrays by sharing the cost.

\subsection{\label{subsec:juna}JUNA: Jinping Underground Nuclear Astrophysics experiment}
\begin{figure}[htbp]
\centering
\includegraphics[width=0.9\linewidth]{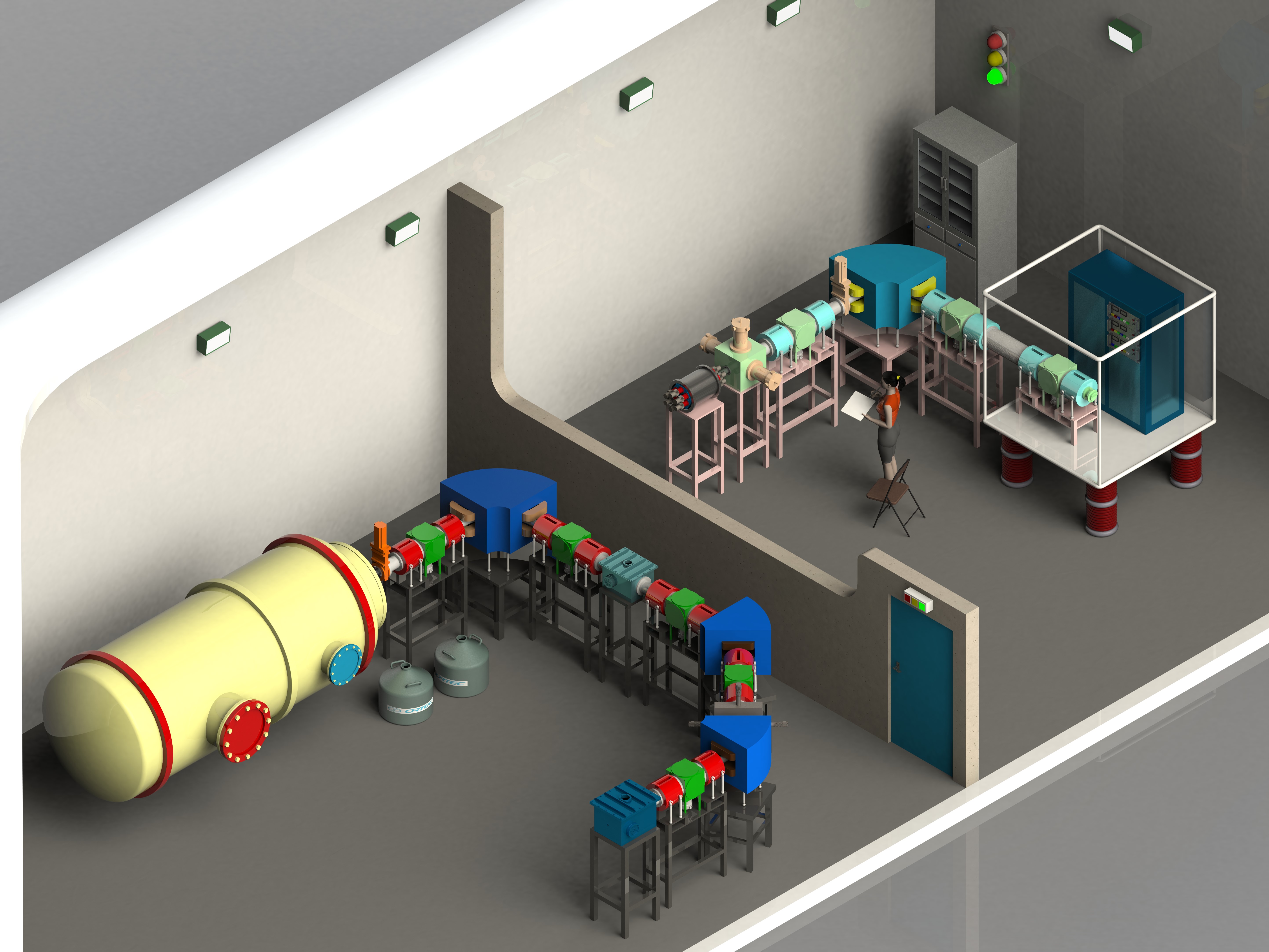}
\caption{The layout of JUNA.}
\label{fig:juna}
\end{figure}

The determination of the very low nuclear reaction rates in stars is a long standing challenge in nuclear astrophysics.
In order to measure the extremely small nuclear cross section in the Gamow window,
one of the most effective methods is to take the advantage of the rock with a thickness of several kilometers to shield the cosmic rays and the corresponding backgrounds. In 2014, the second phase of the China JinPing underground Laboratory (CJPL-II) was initialized. The CJPL-II is 2400 meters deep and listed as the deepest underground laboratory around the world. Jinping Underground lab for Nuclear Astrophysics (JUNA) collaboration is taking the advantage of the ultra-low background of CJPL-II and the high intensity accelerator to measure the critical reactions , $^{12}$C($\alpha$,$\gamma$)$^{16}$O, $^{13}$C($\alpha$,n)$^{16}$O, $^{19}$F(p,$\alpha$)$^{16}$O and $^{25}$Mg(p,$\gamma$)$^{26}$Al at their stellar energies~\cite{JUNA_2016SCPMA}. The layout of JUNA and other CJPL-II projects can be found in Fig.~\ref{fig:juna}.
JUNA is managed by CIAE, jointly supported by NSFC, CNNC and CAS. Development of high intensity accelerator have been completed. It is compatible to deliver proton and $^4$He$^+$ beams with an intensity up to 10 emA and $^4$He$^{2+}$ beam with an intensity up to 2 emA. A number of test experiments have been carried out at the ground level. The first beam was delivered in the Jinping Underground lab in Dec. 2020 and the experimental campaign began in Jan. 2021 to provide more accurate reaction rates for the studies such as s-process, helium burning in stars and the galactic radioactivities. JUNA collaboration is also planning to add a heavy ion accelerator with higher energies to study the critical reactions, such as $^{12}$C+$^{12}$C, in the advanced burning phases of stars.

\section{Conclusion}
By detecting photons with the electromagnetic interaction, neutrinos with the weak interaction, cosmic rays with the strong interaction, GW with the gravitational interaction, multi-messenger astronomy has greatly enriched our ways of perceiving the universe. The successful multi-messenger observation of GW170817 has advanced greatly our knowledge of r-process and the EoS of nuclear matter. The collaborative efforts of LAMOST and Subaru discovered an r-process-enhanced star with relatively high metallicity, suggesting a hierarchical structure formation scenario of the Milky Way.
The progresses made at nuclear facilities, such as discoveries of new isotopes and precise measurements of nuclear mass and lifetime, precise S(E2) factor of $^{12}$C($\alpha$,$\gamma$)$^{16}$O and measurements of the difficult reactions involving $^{7}$Be, have eliminated some critical nuclear uncertainties, and provide a more solid foundation for the development of stellar models. The JUNA collaboration is likely to achieve some breakthroughs in the direct measurements of the critical cross sections at stellar energies with the intense beams in Jinping deep underground lab. New method of producing even more neutron-rich isotopes is being pursued and likely to open a new door towards the r-process studies.  

Future observatories, such as TMT and JUNO, will bring the human eyesight even further in the Universe and deeper inside of stars.  As the new nuclear facilities such as RAON and HIAF and better technologies become available, we will continue to challenge the two long standing problems, the determination of the very low nuclear reaction rates in stars and the determination of properties and reactions of very neutron deficient or very neutron rich unstable nuclei. Complimentary to the other reaction library projects in US and Europe, theoretical collaborations such as A3LIB are developing global databases for the astrophysical applications with predictable errors. With the accurate nuclear reaction rates, more powerful computing resources, and advances in theory aspects, stellar models will become more realistic and reliable and eventually reveal the truths of stars carried by the multi-messengers. 

 Interdisciplinary feature of nuclear astrophysics demands  the close collaborations among astronomers, astrophysicists and nuclear physicists and among the facilities. As we demonstrated in the paper, NO single facility or model will answer all the quests in our field. How to be successful in nuclear astrophysics? Here are the advises from Willy Fowler: seek for truth, work hard and help people.


\section*{Declaration}
\subsection*{Ethical Approval and Consent to participate}
Not applicable
\subsection*{Consent for publication}
Not applicable
\subsection*{Availability of data and materials}
Not applicable
\subsection*{Competing interests}
The authors declare that they have no competing interests.
\subsection*{Funding}
This work is supported in part by the National Key Research and Development program (MOST 2016YFA0400501) from
the Ministry of Science and Technology of China, the Strategic Priority Research Program of Chinese Academy of Sciences (No. XDB34020200) and Grants-in-Aid for Scientific Research of JSPS (20K03958, 17K05459).
For Korea: KIH, SA, TSP and DK are supported by the Institute for Basic Science (IBS-R031-D1).
For Malaysia: HAK, NY and NS are supported in part by MOHE FRGS  FP042-2018A and UMRG GPF044B-2018 grants; AAA by IIUM RIGS2016 grant; MB by  FOS-TECT.2019B.04 and FAPESP 2017/05660-0 grants.
For Taiwan: MRW and GG acknowledge supports from the MOST under Grant No.~109-2112-M-001-004 and
the Academia Sinica under project number AS-CDA-109-M11. KCP is supported by the MOST through grants MOST 107-2112-M-007-032-MY3.
KC was supported by the EACOA Fellowship and by MOST under Grant no. MOST 107-2112-M-001-044-MY3.
\subsection*{Authors' contributions}
All authors contributed equally to all aspects of the manuscript. All authors read and approved the final manuscript. 
\subsection*{Acknowledgements}
The authors appreciate the valuable discussions with Bo Wang, Chengyuan Wu, Z.G. Xiao, Hsien Shang, Hidetoshi Yamaguchi and contributions from Hirokazu Tamura, Tatsushi Shima, Shunji Nishimura, Meng Wang, X.H. Zhou.

%
%

\bibliographystyle{spphys_rev3}       
\bibliography{asia_ref_v3}   

\end{document}